\documentclass[12pt]{article}

\author{Yu.~M.~Zinoviev
       \thanks{E-mail address: Yurii.Zinoviev@ihep.ru} \\
        {\it Institute for High Energy Physics} \\
        {\it Protvino, Moscow Region, 142280, Russia}}
\title{Frame-like gauge invariant formulation\\
        for mixed symmetry fermionic fields}

\date{}

\begin{document}

\setlength{\unitlength}{1mm}

\maketitle

\begin{abstract}
In this paper we consider frame-like formulation for mixed symmetry
spin-tensors corresponding to arbitrary Young tableau with two rows.
First of all, we extend Skvortsov formulation \cite{Skv08} for
massless mixed symmetry bosonic fields in flat Minkowski space to the
case of massless fermionic fields. Then, using such massless fields as
building blocks, we construct gauge invariant formulation for massive
spin-tensors with the same symmetry properties. We give general
massive theories in $(A)dS$ spaces with arbitrary cosmological
constant and investigate all possible massless and partially massless
limits.
\end{abstract}

\thispagestyle{empty}
\newpage
\setcounter{page}{1}

\tableofcontents

\section{Introduction}

As is well known, in $d=4$ dimensions for the description of arbitrary
spin particles it is enough to consider completely symmetric
(spin-)tensor fields only. At the same time, in dimensions greater
than four in many cases like supergravity theories, superstrings and
higher spin theories, one has to deal with mixed symmetry
(spin-)tensor fields \cite{Cur86,AKO86,LM86,Lab89}. There are
different approaches to investigation of such fields both light-cone
\cite{Met02,Met04}, as well as explicitly Lorentz covariant ones (e.g.
\cite{BPT01,BB02,Zin02a,BB06,BKT07,MR07,CFMS08}). For the
investigation of possible interacting theories for higher spin
particles as well as of gauge symmetry algebras behind them it is very
convenient to use  so-called frame-like formulation
\cite{Vas80,LV88,Vas88} (see also \cite{SV06,SV08,Zin08b}) which is a
natural generalization of well-known frame formulation of gravity in
terms of veilbein $e_\mu{}^a$ and Lorentz connection
$\omega_\mu{}^{ab}$.

There are two different frame-like formulations for massless mixed
symmetry bosonic fields. For simplicity, let us restrict ourselves
with mixed symmetry tensors corresponding to Young tableau with two
rows. Let us denote $Y(k,l)$ a tensor $\Phi^{a_1 \dots a_k,b_1 \dots
b_l}$ which is symmetric both on first $k$ as well as last $l$
indices, completely traceless on all indices and satisfies a
constraint $\Phi^{(a_1 \dots a_k,b_1) b_2 \dots b_l} = 0$, where round
brackets mean symmetrization. In the first approach
\cite{ASV03,Alk03,ASV05,ASV06} for the description of $Y(k,l)$ tensor
($k \ne l$) one use a one-form $e_\mu{}^{Y(k-1,l)}$ as a main physical
field. In this, only one of two gauge symmetries is realized
explicitly and such approach is very well adapted for the $AdS$
spaces. Another formulation \cite{Skv08} uses two-form
$e_{\mu\nu}{}^{Y(k-1,l-1)}$ as a main physical field in this, both
gauge symmetries are realized explicitly. Such formalism works in flat
Minkowski space while deformations into $AdS$ space requires
introduction of additional fields \cite{BMV00}.

In Section 2 of our paper we extend the formulation of \cite{Skv08} to
the case of mixed-symmetry spin-tensors 
$Y(k+\frac{1}{2},l+\frac{1}{2})$ corresponding to arbitrary Young
tableau with two rows. Similarly to the bosonic case both gauge
transformations will be realized explicitly and formulation will work
in flat Minkowski space only while deformation into $AdS$ space
turns out to be impossible (the only exception is a spin-tensor
$Y(k+\frac{1}{2},k+\frac{1}{2})$ corresponding to rectangular Young
tableau).

Then in Section 3 we construct gauge invariant frame-like formulation
for massive mixed symmetry spin-tensors corresponding to arbitrary
Young tableau with two rows (examples for bosonic fields were
considered in \cite{Zin08c}). There are two general approaches to
gauge invariant description of massive fields. One of them uses
powerful BRST approach 
\cite{BK05,BKRT06,BKL06,BKR07,BKT07,MR07,BKR08,BKR09}.
Another one, which we will follow in this work,
\cite{KZ97,Zin01,Med03,Met06,Zin08b,Zin08c,BG08}
is a generalization to higher spin fields of well-known mechanism of
spontaneous gauge symmetry breaking. In this, one starts with
appropriate set of massless fields with all their gauge symmetries and
obtain gauge invariant description of massive field as a smooth
deformation. One of the nice feature of gauge invariant formulation
for massive fields is that it allows us effectively use all known
properties of massless fields serving as building blocks.
As we have already seen in all cases considered previously and we will
see again in this paper, gauge invariant description of massive fields
always allows smooth deformation into $(A)dS$ space without
introduction of any additional fields besides those that are necessary
in flat Minkowski space so that restriction mentioned above will not
be essential for us.

As we will see in all models constructed in Section 3, gauge
invariance completely fixes all parameters in the Lagrangian and gauge
transformations leaving us only one free parameter having dimension of
mass. It is hardly possible to give meaningful definition of what is
mass for mixed symmetry (spin)-tensor fields in $(A)dS$ spaces (see
e.g. \cite{Gar03}) and we will not insist on any such definition.
Instead, we will simply use this parameter to analyze all possible
special limits that exist in $(A)dS$ spaces. In this, only fields
having the same number of degrees of freedom as massless one in flat
Minkowski space we will call massless ones, while all other special
limits that appear in $(A)dS$ spaces will be called partially massless
\cite{DW01,DW01a,DW01c,Zin01,SV06}.

\section{Massless case}

In this Section we consider frame-like formulation for massless mixed
symmetry fermionic fields in flat Minkowski space. We begin with some
simple concrete examples and then consider their generalization up to
spin-tensors corresponding to arbitrary Young tableau with two rows.
In all cases we also consider a possibility to deform such theories
into $AdS$ space. As is well known, most of mixed symmetry
(spin)-tensors do not admit such deformation without introduction of
some additional fields \cite{BMV00}, but the structure of possible
mass terms and corresponding corrections to gauge transformations will
be heavily used in the next Section where we consider massive
theories.

\subsection{$Y(k+\frac{3}{2},\frac{1}{2})$}

In what follows we will need frame-like formulation for completely
symmetric spin-tensors \cite{LV88,Vas88,Zin08c}. For completeness we
reproduce here all necessary formulas. Main object --- one-form 
$\Phi_\mu{}^{a_1 \dots a_k} = \Phi_\mu{}^{(a_k)}$ completely symmetric
on local indices and satisfying a constraint $\gamma^{a_1} 
\Phi_\mu{}^{a_1(a_{k-1})} = (\gamma \Phi)_\mu{}^{(a_{k-1})} = 0$.
To describe correct number of physical degrees of freedom the free
massless theory have to be invariant under the following gauge
transformations:
\begin{equation}
\delta_0 \Phi_\mu{}^{(a_k)} = \partial_\mu \zeta^{(a_k)} +
\eta_\mu{}^{(a_k)}
\end{equation}
where parameters $\zeta$ and $\eta$ have to satisfy:
$$
(\gamma \zeta)^{(a_{k-1})} = 0, \qquad
\eta^{(a,a_k)} = 0, \qquad
\gamma^a \eta^{a,(a_k)} = (\gamma \eta)^{a,(a_{k-1})} = 0
$$
Here and in what follows round brackets denote symmetrization. The
free Lagrangian describing massless particle in flat Minkowski space
can be written as follows:
\begin{equation}
{\cal L}_0 = - i (-1)^k \left\{ \phantom{|}^{\mu\nu\alpha}_{abc}
\right\} [ \bar{\Phi}_\mu{}^{(a_k)} \Gamma^{abc} \partial_\nu
\Phi_\alpha{}^{(a_k)} - 6k \bar{\Phi}_\mu{}^{a(a_{k-1})} \gamma^b
\partial_\nu \Phi_\alpha{}^{c(a_{k-1})} ]
\end{equation}
where relative coefficients are fixed by the invariance under $\eta$
shifts. Here and further:
$$
\left\{ \phantom{|}^{\mu\nu\alpha}_{abc} \right\} =
e^{[\mu}{}_a e^\nu{}_b e^{\alpha]}{}_c, \qquad
\Gamma^{abc} = \frac{1}{6} \gamma^{[a} \gamma^b \gamma^{c]}
$$
and so on. It is not hard to construct a deformation into $AdS$ space.
If we replace ordinary partial derivatives in the Lagrangian and gauge
transformations by the $AdS$ covariant ones, the Lagrangian cease to
be gauge invariant:
$$
\delta_0 {\cal L}_0 = i (-1)^{k+1} \frac{(d+2k-1)(d+2k-2)}{2} \kappa
$$
Note that the Lagrangian is completely antisymmetric on world indices,
so covariant derivatives effectively act on local indices (including
implicit spinor one) only, e.g.:
$$
[ D_\mu, D_\nu ] \zeta^{(a_k)} = - \kappa [ e_{[\mu}{}^{(a_1}
\zeta_{\nu]}{}^{a_{k-1})} + \frac{1}{2} \Gamma_{\mu\nu} \zeta^{(a_k)}
], \qquad \kappa = \frac{2 \Lambda}{(d-1)(d-2)}
$$
But gauge invariance could be restored by adding appropriate mass-like
terms to the Lagrangian as well as corresponding corrections to gauge
transformations:
\begin{equation}
{\cal L}_1 = (-1)^k b_k \left\{ \phantom{|}^{\mu\nu}_{ab} \right\} [
\bar{\Phi}_\mu{}^{(a_k)} \Gamma^{ab} \Phi_\nu{}^{(a_k)} + 2k
\bar{\Phi}_\mu{}^{a(a_{k-1})} \Phi_\nu{}^{b(a_{k-1})} ]
\end{equation}
\begin{equation}
\delta_1 \Phi_\mu{}^{(a_k)} = i \beta_k [ \gamma_\mu \zeta^{(a_k)} -
\frac{2}{(d+2k-2)} \gamma^{(a_1} \zeta_\mu{}^{a_{k-1})} ]
\end{equation}
provided:
$$
\beta_k = - \frac{b_k}{3(d-2)}, \qquad
b_k{}^2 = - \frac{9}{4} (d+2k-2)^2 \kappa
$$
Note that relative coefficients in the mass-like terms are again fixed
by the invariance under $\eta$ shifts, while the structure of
variations are chosen so that they are $\gamma$-transverse.

\subsection{$Y(\frac{5}{2},\frac{3}{2})$}

Let us begin with the simplest example of mixed symmetry fermionic
field. Frame-like description requires two-form $\Psi_{\mu\nu}{}^a$
which is $\gamma$-transverse $\gamma^a \Psi_{\mu\nu}{}^a = 0$. Free
massless theory in flat Minkowski space has to be invariant under the
following gauge transformations:
\begin{equation}
\delta_0 \Psi_{\mu\nu}{}^a = \partial_{[\mu} \xi_{\nu]}{}^a +
\eta_{\mu\nu}{}^a
\end{equation}
where parameter $\xi_\mu{}^a$ is $\gamma$-transverse $\gamma^a
\xi_\mu{}^a = 0$, while parameter $\eta^{abc}$ is completely
antisymmetric and $\gamma$-transverse $\gamma ^a \eta^{abc} = 0$. The
Lagrangian can be written in the following form:
\begin{equation}
{\cal L}_0 = i \left\{ \phantom{|}^{\mu\nu\alpha\beta\gamma}_{abcde}
\right\} [ \bar{\Psi}_{\mu\nu}{}^f \Gamma^{abcde} \partial_\alpha
\Psi_{\beta\gamma}{}^f - 10 \bar{\Psi}_{\mu\nu}{}^a \Gamma^{bcd}
\partial_\alpha \Psi_{\beta\gamma}{}^e ]
\end{equation}
Being completely antisymmetric on world indices, both terms are
separately invariant under the $\xi$ transformations, while the
relative coefficients are fixed by the the invariance under the $\eta$
shifts. 

As is well known it is impossible to deform such massless theory into
$AdS$ space without introduction of additional fields. Indeed, after
replacement of ordinary partial derivatives by the $AdS$ covariant
ones, we could try to restore broken gauge invariance by adding
mass-like terms to the Lagrangian and corresponding corrections to
gauge transformations:
\begin{equation}
{\cal L}_1 = a_1 \left\{ \phantom{|}^{\mu\nu\alpha\beta}_{abcd}
\right\} [ \bar{\Psi}_{\mu\nu}{}^e \Gamma^{abcd} 
\Psi_{\alpha\beta}{}^e + 6 \bar{\Psi}_{\mu\nu}{}^a \Gamma^{bc}
\Psi_{\alpha\beta}{}^d ]
\end{equation}
\begin{equation}
\delta \Psi_{\mu\nu}{}^a =  i \alpha_1 [ \gamma_{[\mu} \xi_{\nu]}{}^a
+ \frac{2}{d} \gamma^a \xi_{[\mu,\nu]} ]
\end{equation}
In this, variations with one derivative cancel provided:
$$
\alpha_1 = - \frac{a_1}{5(d-4)}
$$
but it is impossible to cancel variations without derivatives by
adjusting the only free parameter $a_1$.

\subsection{$Y(k+\frac{3}{2},\frac{3}{2})$}

It is pretty straightforward to generalize the example of previous
Subsection to the case corresponding to Young tableau with $k+1$ boxes
in the first row and only one box in the second row. Frame-like
formulation requires two-form $\Psi_{\mu\nu}{}^{(a_k)}$ completely
symmetric on its $k$ local indices and $\gamma$-transverse 
$\gamma^{a_1} \Psi_{\mu\nu}{}^{a_1(a_{k-1})} = 0$. Free massless
theory has to be invariant under the following gauge transformations:
\begin{equation}
\delta_0 \Psi_{\mu\nu}{}^{(a_k)} = \partial_{[\mu} 
\xi_{\nu]}{}^{(a_k)} + \eta_{\mu\nu}{}^{(a_1,a_{k-1})}
\end{equation}
where parameter $\xi_\mu{}^{(a_k)}$ is $\gamma$-transverse, while
parameter $\eta^{abc,(a_{k-1})}$ completely antisymmetric on first
three indices, completely symmetric on the last $k-1$ ones and
satisfies:
$$
\eta^{[abc,a_1](a_{k-2})} = 0, \qquad
\gamma^a \eta^{abc,(a_{k-1})} =  
(\gamma \eta)^{abc,(a_{k-2})} = 0
$$
Corresponding massless Lagrangian has the form:
\begin{equation}
{\cal L}_0 = i (-1)^{k+1} \left\{
\phantom{|}^{\mu\nu\alpha\beta\gamma}_{abcde}
\right\} [ \bar{\Psi}_{\mu\nu}{}^{(a_k)} \Gamma^{abcde}
\partial_\alpha \Psi_{\beta\gamma}{}^{(a_k)} - 10k 
\bar{\Psi}_{\mu\nu}{}^{a(a_{k-1})} \Gamma^{bcd} \partial_\alpha
\Psi_{\beta\gamma}{}^{e(a_{k-1})} ]
\end{equation}

Exactly as in the previous case an attempt to deform such theory into
$AdS$ space without introduction of additional fields fails. Again,
after replacement of ordinary derivatives by the covariant ones, we
could try to restore broken gauge invariance by adding mass-like terms
to the Lagrangian as well as corresponding corrections to gauge
transformations:
\begin{equation}
{\cal L}_1 = (-1)^k a_k \left\{ \phantom{|}^{\mu\nu\alpha\beta}_{abcd}
\right\} [ \bar{\Psi}_{\mu\nu}{}^{(a_k)} \Gamma^{abcd} 
\Psi_{\alpha\beta}{}^{(a_k)} + 6k \bar{\Psi}_{\mu\nu}{}^{a(a_{k-1})}
\Gamma^{bc} \Psi_{\alpha\beta}{}^{d(a_{k-1})} ]
\end{equation}
\begin{equation}
\delta_1 \Psi_{\mu\nu}{}^{(a_k)} = i \alpha_k [ \gamma_{[\mu} 
\xi_{\nu]}{}^{(a_k)} + \frac{2}{(d+2k-2)} \gamma^{(a_1} 
\xi_{[\mu,\nu]}{}^{a_{k-1})} ]
\end{equation}
In this, variations with one derivative cancel provided:
$$
\alpha_k = \frac{a_k}{5(d-4)}
$$
but it is impossible to achieve the cancellation of variations without
derivatives.

\subsection{$Y(\frac{5}{2},\frac{5}{2})$}

Among all mixed symmetry (spin)-tensors corresponding to Young tableau
with two rows whose with equal number of boxes in both rows turn out
to be special and require separate consideration. Let us begin with
the simplest example --- $Y(\frac{5}{2},\frac{5}{2})$. Frame-like
formulation requires two-form $R_{\mu\nu}{}^{ab}$ which is
antisymmetric on $ab$ and $\gamma$-transverse $\gamma^a
R_{\mu\nu}{}^{ab} = 0$. Free massless theory has to be invariant under
the following gauge transformations:
\begin{equation}
\delta_0 R_{\mu\nu}{}^{ab} = \partial_{[\mu} \xi_{\nu]}{}^{ab} + 
\eta_{[\mu,\nu]}{}^{ab}
\end{equation}
where parameters $\xi_\mu{}^{ab}$ and $\eta_\mu{}^{abc}$ are
antisymmetric on their local indices and $\gamma$-transverse. It is
not hard to construct gauge invariant Lagrangian:
\begin{equation}
{\cal L}_0 = - i \left\{ \phantom{|}^{\mu\nu\alpha\beta\gamma}_{abcde}
\right\} [ \bar{R}_{\mu\nu}{}^{fg} \Gamma^{abcde} \partial_\alpha
R_{\beta\gamma}{}^{fg} - 20 \bar{R}_{\mu\nu}{}^{af} \Gamma^{bcd}
\partial_\alpha R_{\beta\gamma}{}^{ef} - 60 \bar{R}_{\mu\nu}{}^{ab}
\gamma^c \partial_\alpha R_{\beta\gamma}{}^{de} ]
\end{equation}
where again each term is separately invariant under the $\xi$
transformations, while relative coefficients are fixed by the
invariance under the $\eta$ shifts.

One of the main special features of such (spin)-tensors is the fact
that they admit deformation into $AdS$ space without introduction of
any additional fields. Indeed, let us replace all derivatives in the
Lagrangian and gauge transformations by the $AdS$ covariant ones. As
usual, the initial Lagrangian cease to be invariant:
$$
\delta_0 {\cal L}_0 = - 10 i (d-1)(d-2) \kappa
\left\{ \phantom{|}^{\mu\nu\alpha}_{abc} \right\} [
\bar{R}_{\mu\nu}{}^{de} \Gamma^{abc} \xi_\alpha{}^{de} - 6
\bar{R}_{\mu\nu}{}^{ad} \gamma^b \xi_\alpha{}^{cd} ]
$$
In this, broken gauge invariance can be restored by adding mass-like
terms to the Lagrangian and corresponding corrections to gauge
transformations:
\begin{equation}
{\cal L}_1 = a_{1,1} \left\{ \phantom{|}^{\mu\nu\alpha\beta}_{abcd}
\right\} [ \bar{R}_{\mu\nu}{}^{ef} \Gamma^{abcd} 
R_{\alpha\beta}{}^{ef} + 12 \bar{R}_{\mu\nu}{}^{ae} \Gamma^{bc}
R_{\alpha\beta}{}^{de} - 12 \bar{R}_{\mu\nu}{}^{ab} 
R_{\alpha\beta}{}^{cd} ]
\end{equation}
\begin{equation}
\delta_1 R_{\mu\nu}{}^{ab} = i \alpha_{1,1} [ \gamma_{[\mu} 
\xi_{\nu]}{}^{ab} + \frac{2}{(d-2)} \gamma^{[a} \xi_{[\mu,\nu]}{}^{b]}
]
\end{equation}
provided:
$$
\alpha_{1,1} = \frac{a_{1,1}}{5(d-4)}, \qquad
a_{1,1}{}^2 = - \frac{25(d-2)^2}{4} \kappa
$$

\subsection{$Y(k+\frac{3}{2},k+\frac{3}{2})$}

It is straightforward to construct a generalization of previous
example for arbitrary $k > 1$. For this we need a two-form 
$R_{\mu\nu}{}^{(a_k),(b_k)}$ which is symmetric and
$\gamma$-transverse on both groups of local indices and satisfies
$R_{\mu\nu}{}^{(a_k,b_1)(b_{k-1})} = 0$.  Moreover,
$R_{\mu\nu}{}^{(a_k),(b_k)} = R_{\mu\nu}{}^{(b_k),(a_k)}$. Free
massless theory has to be invariant under the following gauge
transformations:
\begin{equation}
\delta_0 R_{\mu\nu}{}^{(a_k),(b_k)} = D_{[\mu} 
\xi_{\nu]}{}^{(a_k),(b_k)} + \eta_{[\mu}{}^{(a_k),(b_k)}{}_{\nu]}
\end{equation}
where parameter $\xi_\mu{}^{(a_k),(b_k)}$ has the same properties on
local indices as $R_{\mu\nu}{}^{(a_k),(b_k)}$, while parameter 
$\eta^{(a_k),(b_k),a}$ satisfies:
$$
\eta_\mu{}^{(a_k),(b_k,a)} = 0, \qquad
(\gamma \eta)_\mu{}^{(a_{k-1}),(b_k),a} = 
(\gamma \eta)_\mu{}^{(a_k),(b_{k-1}),a} = 
\gamma^a \eta_\mu{}^{(a_k),(b_k),a} = 0
$$
Massless Lagrangian can be constructed out of three terms separately
invariant under $\xi_\mu$ transformations:
\begin{eqnarray}
{\cal L}_0 &=&- i \left\{
\phantom{|}^{\mu\nu\alpha\beta\gamma}_{abcde} \right\} [
\bar{R}_{\mu\nu}{}^{(a_k),(b_k)} \Gamma^{abcde} \partial_\alpha
R_{\beta\gamma}{}^{(a_k),(b_k)} - 20 k 
\bar{R}_{\mu\nu}{}^{(a_k),a(b_{k-1})} \Gamma^{bcd} \partial_\alpha
R_{\beta\gamma}{}^{(a_k),e(b_{k-1})} + \nonumber \\
 && \qquad \qquad \quad - 60 k^2
\bar{R}_{\mu\nu}{}^{a(a_{k-1}),b(b_{k-1})} \gamma^c \partial_\alpha
R_{\beta\gamma}{}^{d(a_{k-1}),e(b_{k-1})} ] \label{lagkk0}
\end{eqnarray}
where relative coefficients are fixed by the invariance under $\eta$
transformations.

As in the previous case, such massless theory could be deformed into
$AdS$ space without introduction of any additional fields. Gauge
invariance broken by the replacement of ordinary derivatives by the
$AdS$ covariant ones can be restored if we add to the Lagrangian
mass-like terms of the form:
\begin{eqnarray}
{\cal L}_1 &=& a_{k,k} \left\{ \phantom{|}^{\mu\nu\alpha\beta}_{abcd}
\right\} [ \bar{R}_{\mu\nu}{}^{(a_k),(b_k)} \Gamma^{abcd} 
R_{\alpha\beta}{}^{(a_k),(b_k)} + 12 k
\bar{R}_{\mu\nu}{}^{(a_k),a(b_{k-1})} \Gamma^{bc}
R_{\alpha\beta}{}^{(a_k),d(b_{k-1})} + \nonumber \\
 && \qquad \qquad \quad - 12 k^2 
\bar{R}_{\mu\nu}{}^{a(a_{k-1}),b(b_{k-1})}
R_{\alpha\beta}{}^{c(a_{k-1}),d(b_{k-1})} ] \label{lagkk1}
\end{eqnarray}
as well as corresponding corrections to gauge transformations:
\begin{equation}
\delta_1 R_{\mu\nu}{}^{(k),(k)} = i \alpha_{k,k} [ \gamma_{[\mu}
\xi_{\nu]}{}^{(a_k),(b_k)} + \frac{2}{(d+2k-4)} ( \gamma^{(a_1}
\xi_{[\mu,\nu]}{}^{a_{k-1}),(b_k)} + \gamma^{(b_1} 
\xi_{[\mu}{}^{(a_k),b_{k-1})}{}_{\nu]} ) ]
\end{equation}
provided:
$$
\alpha_{k,k} = \frac{a_{k,k}}{5(d-4)}, \qquad
a_{k,k}{}^2 = - \frac{25}{4} (d+2k-4)^2 \kappa
$$

\subsection{$Y(k+\frac{3}{2},l+\frac{3}{2})$}

Now we are ready to consider general case of 
$Y(k+\frac{3}{2},l+\frac{3}{2})$ with $k > l \ge 1$. This time we need
a two-form  $\Psi_{\mu\nu}{}^{(a_k),(b_l)}$ which is symmetric and
$\gamma$-transverse on both groups of local indices and satisfies
$\Psi_{\mu\nu}{}^{(a_k,b_1)(b_{l-1})} = 0$. Gauge transformations for
free massless theory have the form:
\begin{equation}
\delta \Psi_{\mu\nu}{}^{(a_k),(b_l)} = D_{[\mu} 
\xi_{\nu]}{}^{(a_k),(b_l)} + \eta_{[\mu}{}^{(a_k),(b_l)}{}_{\nu]}
\end{equation}
where parameter $\xi_\mu{}^{(a_k),(b_l)}$ has the same properties on
local indices as $\Psi_{\mu\nu}{}^{(a_k),(b_l)}$, while parameter
$\eta^{(a_k),(b_l),c}$ satisfies:
$$
\eta^{(a_k,b_1)(b_{l-1}),c} =
\eta_\mu{}^{(a_k),(b_l,c)} = 0, \qquad
(\gamma \eta)_\mu{}^{(a_{k-1}),(b_l),c} = 
(\gamma \eta)_\mu{}^{(a_k),(b_{l-1}),c} = 
\gamma^c \eta_\mu{}^{(a_k),(b_l),c} = 0
$$
This time we have four terms separately invariant under $\xi_\mu$
transformations to construct massless Lagrangian:
\begin{eqnarray}
i (-1)^{k+l} {\cal L}_0 &=&  
\left\{ \phantom{|}^{\mu\nu\alpha\beta\gamma}_{abcde}
\right\} [ \bar{\Psi}_{\mu\nu}{}^{(a_k),(b_l)} \Gamma^{abcde}
\partial_\alpha \Psi_{\beta\gamma}{}^{(a_k),(b_l)} - \nonumber \\
 && \qquad \qquad - 10 k
\bar{\Psi}_{\mu\nu}{}^{a(a_{k-1}),(b_l)} \Gamma^{bcd} \partial_\alpha
\Psi_{\beta\gamma}{}^{e(a_{k-1}),(b_l)} + \nonumber \\
 && \qquad \qquad - 10 l 
\bar{\Psi}_{\mu\nu}{}^{(a_k),a(b_{l-1})} \Gamma^{bcd} \partial_\alpha
\Psi_{\beta\gamma}{}^{(a_k),e(b_{l-1})} - \nonumber \\
 && \qquad \qquad -
60 kl \bar{\Psi}_{\mu\nu}{}^{a(a_{k-1}),b(b_{l-1})} \gamma^c
\partial_\alpha \Psi_{\beta\gamma}{}^{d(a_{k-1}),e(b_{l-1})} ]
\label{lagkl0}
\end{eqnarray}
where as usual relative coefficients are fixed by the invariance under
$\eta$ transformations.

It is not possible to deform this massless theory into $AdS$ space
without introduction of additional fields. Indeed, possible mass-like
terms look as follows:
\begin{eqnarray}
(-1)^{k+l} {\cal L}_1 &=&  a_{k,l}
\left\{ \phantom{|}^{\mu\nu\alpha\beta}_{abcd} \right\} [
\bar{\Psi}_{\mu\nu}{}^{(a_k),(b_l)} \Gamma^{abcd} 
\Psi_{\alpha\beta}{}^{(a_k),(b_l)} + \nonumber \\
 && \qquad \qquad \quad + 6 l
\bar{\Psi}_{\mu\nu}{}^{(a_k),a(b_{l-1})} \Gamma^{bc}
\Psi_{\alpha\beta}{}^{(a_k),d(b_{l-1})} + \nonumber \\
 && \qquad \qquad \quad + 6 k
\bar{\Psi}_{\mu\nu}{}^{a(a_{k-1}),(b_l)} \Gamma^{bc}
\Psi_{\alpha\beta}{}^{d(a_{k-1}),(b_l)} - \nonumber \\
 && \qquad \qquad \quad - 12 kl
\bar{\Psi}_{\mu\nu}{}^{a(a_{k-1}),b(b_{l-1})}
\Psi_{\alpha\beta}{}^{c(a_{k-1}),d(b_{l-1})} ] \label{lagkl1}
\end{eqnarray}
In this, their non-invariance under the initial gauge transformations
can be compensated by corresponding corrections to gauge
transformations:
\begin{eqnarray}
\delta_1 \Psi_{\mu\nu}{}^{(a_k),(b_l)} &=& i \alpha_{k,l} [
\gamma_{[\mu} \xi_{\nu]}{}^{(a_k),(b_l)} + \frac{2}{(d+2k-2)}
\gamma^{(a_1} \xi_{[\mu,\nu]}{}^{a_{k-1}),(b_l)} + \\
 && + \frac{2}{(d+2l-4)} \gamma^{(b_1} 
\xi_{[\mu}{}^{(a_k),b_{l-1})}{}_{\nu]} - \frac{4}{(d+2k-2)(d+2l-4)}
\gamma^{(a_1} \xi_{[\mu}{}^{a_{k_1})(b_1,b_{l-1})}{}_{\nu]}] \nonumber
\end{eqnarray}
provided:
$$
\alpha_{k,l} = \frac{a_{k,l}}{5(d-4)}
$$
but it is not possible to cancel variations without derivatives by
adjusting the value of $a_{k,l}$.

\section{Massive case}

In this Section we construct gauge invariant frame-like formulation
for massive mixed symmetry fermionic fields. Once again we begin with
some simple concrete examples and then construct their
generalizations. In all cases our general strategy will be the same.
First of all we determine a set of massless fields which are necessary
for gauge invariant description of massive field. Then we construct
the Lagrangian as a sum of kinetic and mass terms for all fields
involved as well as all possible cross terms without derivatives and
look for the necessary corrections to gauge transformations. As we
have already mentioned in the Introduction, such gauge invariant
formalism works equally well both in flat Minkowski space as well as
in $(A)dS$ space with arbitrary value of cosmological constant. This,
in turn, allows us to investigate all possible massless and partially
massless limits that exist in $(A)dS$ spaces.

\subsection{$Y(\frac{5}{2},\frac{3}{2})$}

Let us begin with the simplest case --- $Y(\frac{5}{2},\frac{3}{2})$.
To construct gauge invariant description of massive particle we, first
of all, have to determine a set of massless fields which are necessary
for such description. In general, for each gauge invariance of main
gauge field we have to introduce corresponding primary Goldstone
field. Usually, these fields turn out to be gauge fields themselves
with their own gauge invariances, so we have to introduce secondary
Goldstone fields and so on. But in the mixed symmetry (spin)-tensor
case we have to take into account reducibility of their gauge
transformations. Let us illustrate on this simplest case. Our main
gauge field $Y(\frac{5}{2},\frac{3}{2})$ has two gauge transformations
(combined into one $\xi_\mu{}^a$ transformation in the frame-like
approach) with the parameters $Y(\frac{5}{2},\frac{1}{2})$ and
$Y(\frac{3}{2},\frac{3}{2})$ and reducibility corresponding to
$Y(\frac{3}{2},\frac{1}{2})$. Thus we have to introduce two primary
Goldstone fields corresponding to $Y(\frac{5}{2},\frac{1}{2})$ and
$Y(\frac{3}{2},\frac{3}{2})$. Both have its own gauge transformations
with parameters $Y(\frac{3}{2},\frac{1}{2})$ but due to reducibility
of gauge transformations for the main gauge field, it is enough to
introduce one secondary Goldstone field $Y(\frac{3}{2},\frac{1}{2})$
only. This field also has its own gauge transformation with parameter
$Y(\frac{1}{2},\frac{1}{2})$ but due to reducibility of gauge
transformations for the field $Y(\frac{3}{2},\frac{3}{2})$ the
procedure stops here. Thus we need four fields:
$Y(\frac{5}{2},\frac{3}{2})$, $Y(\frac{5}{2},\frac{1}{2})$,
$Y(\frac{3}{2},\frac{3}{2})$ and $Y(\frac{3}{2},\frac{1}{2})$. It is
natural to use frame-like formalism for all fields in question so we
will use $\Psi_{\mu\nu}{}^a$, $\Phi_\mu{}^a$, $\Psi_{\mu\nu}$ and
$\Phi_\mu$ respectively.

In general, gauge invariant Lagrangian for massive fermionic field
contains kinetic and mass terms for all the components as well as a
number of cross terms without derivatives. Moreover, it is necessary
to introduce such cross terms for the nearest neighbours only, i.e.
main gauge field with the primary ones, primary with secondary and so
on. Thus we will look for gauge invariant Lagrangian in the form:
\begin{eqnarray}
{\cal L} &=& i \left\{ \phantom{|}^{\mu\nu\alpha\beta\gamma}_{abcde}
\right\} [ \bar{\Psi}_{\mu\nu}{}^f \Gamma^{abcde} D_\alpha
\Psi_{\beta\gamma}{}^f - 10 \bar{\Psi}_{\mu\nu}{}^a \Gamma^{bcd}
D_\alpha \Psi_{\beta\gamma}{}^e - \bar{\Psi}_{\mu\nu} \Gamma^{abcde}
D_\alpha \Psi_{\beta\gamma} ] + \nonumber \\
 && + i \left\{ \phantom{|}^{\mu\nu\alpha}_{abc} \right\} [
\bar{\Phi}_\mu{}^d \Gamma^{abc} D_\nu \Phi_\alpha{}^d - 6
\bar{\Phi}_\mu{}^a \gamma^b D_\nu \Phi_\alpha{}^c -
\bar{\Phi}_\mu \Gamma^{abc} D_\nu \Phi_\alpha ] + \nonumber \\
 && + \left\{ \phantom{|}^{\mu\nu\alpha\beta}_{abcd} \right\} [
a_0 \bar{\Psi}_{\mu\nu}{}^e \Gamma^{abcd} \Psi_{\alpha\beta}{}^e + 6
a_0 \bar{\Psi}_{\mu\nu}{}^a \Gamma^{bc} \Psi_{\alpha\beta}{}^d + 
a_2 \bar{\Psi}_{\mu\nu} \Gamma^{abcd} \Psi_{\alpha\beta} ] + \nonumber
\\
 && + \left\{ \phantom{|}^{\mu\nu}_{ab} \right\} [
a_1 ( \bar{\Phi}_\mu{}^c \Gamma^{ab} \Phi_\nu{}^c + 2
\bar{\Phi}_\mu{}^a \Phi_\nu{}^b )  + a_3 \bar{\Phi}_\mu \Gamma^{ab}
\Phi_\nu ] + \nonumber \\
 && + i b_1 \left\{ \phantom{|}^{\mu\nu\alpha}_{abc} \right\} [
\bar{\Psi}_{\mu\nu}{}^d \Gamma^{abc} \Phi_\alpha{}^d - 6
\bar{\Psi}_{\mu\nu}{}^a \gamma^b \Phi_\alpha{}^c + 
\bar{\Phi}_\mu{}^d \Gamma^{abc} \Psi_{\nu\alpha}{}^d - 6
\bar{\Phi}_\mu{}^a \gamma^b \Psi_{\nu\alpha}{}^c ] + \nonumber \\
 && + i b_2 \left\{ \phantom{|}^{\mu\nu\alpha\beta}_{abcd} \right\}
[ \bar{\Psi}_{\mu\nu}{}^a \Gamma^{bcd}
\Psi_{\alpha\beta} -  \bar{\Psi}_{\mu\nu} \Gamma^{abc}
\Psi_{\alpha\beta}{}^d ] + \nonumber \\
 && + i b_3 \left\{ \phantom{|}^{\mu\nu}_{ab} \right\} [
\bar{\Phi}_\mu{}^a \gamma^b \Phi_\nu -
\bar{\Phi}_\mu \gamma^a \Phi_\nu{}^b ]
+ i b_4\left\{ \phantom{|}^{\mu\nu\alpha}_{abc} \right\} [
\bar{\Psi}_{\mu\nu} \Gamma^{abc} \Phi_\alpha +
\bar{\Phi}_\mu \Gamma^{abc} \Psi_{\nu\alpha} ]
\end{eqnarray}
where all derivatives are $AdS$ covariant ones. In order to compensate
the non-invariance of these mass and cross terms under the initial
gauge transformations we have to introduce corresponding corrections
to gauge transformations. And indeed all variations with one
derivative cancel with the following form of gauge transformations:
\begin{eqnarray}
\delta \Psi_{\mu\nu}{}^a &=& D_{[\mu} \xi_{\nu]}{}^a - 
\frac{i a_0}{5(d-4)} [ \gamma_{[\mu} \xi_{\nu]}{}^a + \frac{2}{d}
\gamma^a \xi_{[\mu,\nu]} ]  - \frac{b_2}{10(d-2)} [ e_{[\mu}{}^a
\xi_{\nu]} - \frac{1}{d} \gamma^a \gamma_{[\mu} \xi_{\nu]} ] \nonumber
\\
 && + \frac{b_1}{15(d-3)(d-4)}
 [ \Gamma_{\mu\nu} \zeta^a - \frac{(d^2-7d+16)}{4(d-2)}
 e_{[\mu}{}^a \zeta_{\nu]} + \frac{(d+1)}{4(d-2)} \gamma^a
\gamma_{[\mu} \zeta_{\nu]} ] \nonumber \\
\delta \Phi_\mu{}^a &=& D_\mu \zeta^a - 2 b_1 \xi_\mu{}^a + 
\frac{i a_1}{3(d-2)} [ \gamma_\mu \zeta^a - \frac{2}{d} \gamma^a
\zeta_\mu ] + \frac{b_3}{6(d-1)} [ e_\mu{}^a \zeta - \frac{1}{d}
\gamma^a \gamma_\mu \zeta ] \nonumber \\
\delta \Psi_{\mu\nu} &=& D_{[\mu} \xi_{\nu]} + \frac{b_2}{20}
\xi_{[\mu,\nu]} + \frac{i a_2}{5(d-4)} \gamma_{[\mu} \xi_{\nu]} -
\frac{b_4}{20(d-3)(d-4)} \Gamma_{\mu\nu} \zeta \\
\delta \Phi_\mu &=& D_\mu \zeta + \frac{b_3}{6} \zeta_\mu + 2 b_4
\xi_\mu  - \frac{i a_3}{3(d-2)} \gamma_\mu \zeta \nonumber
\end{eqnarray}
where all coefficients are expressed in terms of Lagrangian parameters
$a_{1,2,3,4}$ and $b_{1,2,3,4}$. Now we calculate all variations
without derivatives (including contribution of kinetic terms due to
non-commutativity of covariant derivatives) and require their
cancellation. This gives us:
$$
a_1 = - \frac{3(d-2)}{5(d-4)} a_0,\qquad
a_2 = - \frac{(d+2)}{d} a_0, \qquad
a_3 = \frac{3(d^2-4)}{5d(d-4)} a_0
$$
$$
a_0{}^2 = \frac{5(d-4)}{3(d-3)} b_1{}^2 - \frac{25}{4} (d-4)^2 \kappa,
$$
$$
20(d+1)(d-4) b_1{}^2 - 3d(d-3) b_2{}^2 = - 600(d+1)(d-2)(d-3)
\kappa
$$
$$
b_3{}^2 = \frac{9(d-1)}{50(d-2)} b_2{}^2, \qquad
b_4{}^2 = \frac{2(d-1)}{(d-2)} b_1{}^2
$$
Let us analyze the results obtained. First of all recall that there is
no strict definition of what is mass in $(A)dS$ spaces. Working with
gauge invariant description of massive particles it is natural to
define massless limit as the one where all Goldstone fields decouple
from the main gauge one. For the case at hands, such a limit requires
that both $b_1 \to 0$ and $b_2 \to 0$ simultaneously. As the third
relation above clearly shows such a limit is possible in flat
Minkowski space $\kappa = 0$ only. For the non-zero values of
cosmological constant we obtain one of the so called partially
massless limits (depending on the sign of $\kappa$). To clarify
subsequent discussion, let us give here a Figure 1 illustrating the
roles of cross terms $b_{1,2,3,4}$.
\begin{figure}[htb]
\begin{center}
\begin{picture}(70,32)
\put(10,21){\framebox(10,10)[]{$\Psi_{\mu\nu}{}^a$}}
\put(30,21){\framebox(10,10)[]{$\Psi_{\mu\nu}$}}
\put(30,1){\framebox(10,10)[]{$\Phi_\mu{}^a$}}
\put(50,1){\framebox(10,10)[]{$\Phi_\mu$}}
\put(20,26){\vector(1,0){10}}
\put(22,26){\makebox(6,6)[]{$b_2$}}
\put(20,21){\vector(1,-1){10}}
\put(26,14){\makebox(6,6)[]{$b_1$}}
\put(40,21){\vector(1,-1){10}}
\put(46,14){\makebox(6,6)[]{$b_4$}}
\put(40,6){\vector(1,0){10}}
\put(42,6){\makebox(6,6)[]{$b_3$}}
\end{picture}
\end{center}
\caption{General massive theory for $Y(\frac{5}{2},\frac{3}{2})$ 
spin-tensor}
\end{figure}
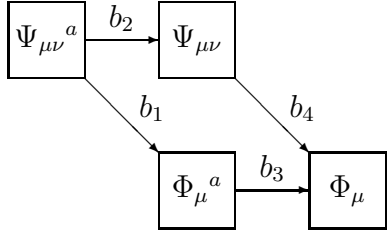
In $AdS$ space ($\kappa < 0$) one can put $b_2 = 0$ (and thus $b_3 =
0$). In this, the whole system decomposes into two disconnected
subsystems as Figure 2 shows.
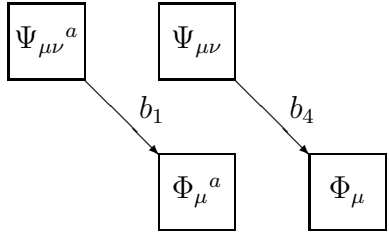
\begin{figure}[htb]
\begin{center}
\begin{picture}(70,32)
\put(10,21){\framebox(10,10)[]{$\Psi_{\mu\nu}{}^a$}}
\put(30,21){\framebox(10,10)[]{$\Psi_{\mu\nu}$}}
\put(30,1){\framebox(10,10)[]{$\Phi_\mu{}^a$}}
\put(50,1){\framebox(10,10)[]{$\Phi_\mu$}}
\put(20,21){\vector(1,-1){10}}
\put(26,14){\makebox(6,6)[]{$b_1$}}
\put(40,21){\vector(1,-1){10}}
\put(46,14){\makebox(6,6)[]{$b_4$}}
\end{picture}
\end{center}
\caption{Partially massless limit for $Y(\frac{5}{2},\frac{3}{2})$
spin-tensor in $AdS$}
\end{figure}
One of them, with the fields $\Psi_{\mu\nu}{}^a$, $\Phi_\mu{}^a$ and
with the Lagrangian:
\begin{eqnarray}
{\cal L} &=& i \left\{ \phantom{|}^{\mu\nu\alpha\beta\gamma}_{abcde}
\right\} [ \bar{\Psi}_{\mu\nu}{}^f \Gamma^{abcde} D_\alpha
\Psi_{\beta\gamma}{}^f - 10 \bar{\Psi}_{\mu\nu}{}^a \Gamma^{bcd}
D_\alpha \Psi_{\beta\gamma}{}^e ] + \nonumber \\
 && + i \left\{ \phantom{|}^{\mu\nu\alpha}_{abc} \right\} [
\bar{\Phi}_\mu{}^d \Gamma^{abc} D_\nu \Phi_\alpha{}^d - 6
\bar{\Phi}_\mu{}^a \gamma^b D_\nu \Phi_\alpha{}^c  ] + \nonumber \\
 && + a_0 \left\{ \phantom{|}^{\mu\nu\alpha\beta}_{abcd} \right\} [
\bar{\Psi}_{\mu\nu}{}^e \Gamma^{abcd} \Psi_{\alpha\beta}{}^e + 6
\bar{\Psi}_{\mu\nu}{}^a \Gamma^{bc} \Psi_{\alpha\beta}{}^d ]
+ a_1 \left\{ \phantom{|}^{\mu\nu}_{ab} \right\} [
 \bar{\Phi}_\mu{}^c \Gamma^{ab} \Phi_\nu{}^c + 2
\bar{\Phi}_\mu{}^a \Phi_\nu{}^b ] + \nonumber \\
 && + i b_1 \left\{ \phantom{|}^{\mu\nu\alpha}_{abc} \right\} [
\bar{\Psi}_{\mu\nu}{}^d \Gamma^{abc} \Phi_\alpha{}^d - 6
\bar{\Psi}_{\mu\nu}{}^a \gamma^b \Phi_\alpha{}^c + 
\bar{\Phi}_\mu{}^d \Gamma^{abc} \Psi_{\nu\alpha}{}^d - 6
\bar{\Phi}_\mu{}^a \gamma^b \Psi_{\nu\alpha}{}^c ]
\end{eqnarray}
which is invariant under the following gauge transformations:
\begin{eqnarray}
\delta \Psi_{\mu\nu}{}^a &=& D_{[\mu} \xi_{\nu]}{}^a - 
\frac{i a_0}{5(d-4)} [ \gamma_{[\mu} \xi_{\nu]}{}^a + \frac{2}{d}
\gamma^a \xi_{[\mu,\nu]} ] \nonumber
\\
 && + \frac{b_1}{15(d-3)(d-4)}
 [ \Gamma_{\mu\nu} \zeta^a - \frac{(d^2-7d+16)}{4(d-2)}
 e_{[\mu}{}^a \zeta_{\nu]} + \frac{(d+1)}{4(d-2)} \gamma^a
\gamma_{[\mu} \zeta_{\nu]} ]  \\
\delta \Phi_\mu{}^a &=& D_\mu \zeta^a - 2 b_1 \xi_\mu{}^a + 
\frac{i a_1}{3(d-2)} [ \gamma_\mu \zeta^a - \frac{2}{d} \gamma^a
\zeta_\mu ]  \nonumber
\end{eqnarray}
describes unitary partially massless theory corresponding to
irreducible representation of $AdS$ group. At the same time, two other
fields $\Psi_{\mu\nu}$ and $\Phi_\mu$ with the Lagrangian:
\begin{eqnarray}
{\cal L} &=& -  i \left\{
\phantom{|}^{\mu\nu\alpha\beta\gamma}_{abcde}
\right\} \bar{\Psi}_{\mu\nu} \Gamma^{abcde}
D_\alpha \Psi_{\beta\gamma} -  i \left\{
\phantom{|}^{\mu\nu\alpha}_{abc} \right\} \bar{\Phi}_\mu \Gamma^{abc}
D_\nu \Phi_\alpha + \nonumber \\
 && + a_2  \left\{ \phantom{|}^{\mu\nu\alpha\beta}_{abcd} \right\}
\bar{\Psi}_{\mu\nu} \Gamma^{abcd} \Psi_{\alpha\beta} + a_3
\left\{ \phantom{|}^{\mu\nu}_{ab} \right\} \bar{\Phi}_\mu \Gamma^{ab}
\Phi_\nu  + \nonumber \\
 && +  i b_4\left\{ \phantom{|}^{\mu\nu\alpha}_{abc} \right\} [
\bar{\Psi}_{\mu\nu} \Gamma^{abc} \Phi_\alpha +
\bar{\Phi}_\mu \Gamma^{abc} \Psi_{\nu\alpha} ]
\end{eqnarray}
invariant under the following gauge transformations:
\begin{eqnarray}
\delta \Psi_{\mu\nu} &=& D_{[\mu} \xi_{\nu]} 
+ \frac{i a_2}{5(d-4)} \gamma_{[\mu} \xi_{\nu]} -
\frac{b_4}{20(d-3)(d-4)} \Gamma_{\mu\nu} \zeta \nonumber \\
\delta \Phi_\mu &=& D_\mu \zeta + 2 b_4
\xi_\mu  - \frac{i a_3}{3(d-2)} \gamma_\mu \zeta 
\end{eqnarray}
give gauge invariant description of massive antisymmetric second rank
spin-tensor \cite{Zin09a}.

Let us turn to the $dS$ space ($\kappa > 0$). First of all, from the
equation for the $a_0{}^2$ above we see that there is a unitary
forbidden region $b_1{}^2 < \frac{15}{4} (d-3)(d-4)\kappa$. Inside
this region "lives" one more example of partially massless theory
corresponding to the limit $b_1 \to 0$ (and hence $b_4 \to 0$) as
Figure 3 shows.
\begin{figure}[htb]
\begin{center}
\begin{picture}(70,32)
\put(10,21){\framebox(10,10)[]{$\Psi_{\mu\nu}{}^a$}}
\put(30,21){\framebox(10,10)[]{$\Psi_{\mu\nu}$}}
\put(30,1){\framebox(10,10)[]{$\Phi_\mu{}^a$}}
\put(50,1){\framebox(10,10)[]{$\Phi_\mu$}}
\put(20,26){\vector(1,0){10}}
\put(22,26){\makebox(6,6)[]{$b_2$}}
\put(40,6){\vector(1,0){10}}
\put(42,6){\makebox(6,6)[]{$b_3$}}
\end{picture}
\end{center}
\caption{Partially massless limit for $Y(\frac{5}{2},\frac{3}{2})$
spin-tensor in $dS$ space}
\end{figure}
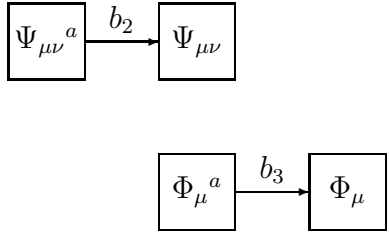
In this, the main field $\Psi_{\mu\nu}{}^a$ together with
$\Psi_{\mu\nu}$ describe this non-unitary partially massless theory
with the Lagrangian:
\begin{eqnarray}
{\cal L} &=& i \left\{ \phantom{|}^{\mu\nu\alpha\beta\gamma}_{abcde}
\right\} [ \bar{\Psi}_{\mu\nu}{}^f \Gamma^{abcde} D_\alpha
\Psi_{\beta\gamma}{}^f - 10 \bar{\Psi}_{\mu\nu}{}^a \Gamma^{bcd}
D_\alpha \Psi_{\beta\gamma}{}^e - \bar{\Psi}_{\mu\nu} \Gamma^{abcde}
D_\alpha \Psi_{\beta\gamma} ] + \nonumber \\
 && + \left\{ \phantom{|}^{\mu\nu\alpha\beta}_{abcd} \right\} [
a_0 \bar{\Psi}_{\mu\nu}{}^e \Gamma^{abcd} \Psi_{\alpha\beta}{}^e + 6
a_0 \bar{\Psi}_{\mu\nu}{}^a \Gamma^{bc} \Psi_{\alpha\beta}{}^d + 
a_2 \bar{\Psi}_{\mu\nu} \Gamma^{abcd} \Psi_{\alpha\beta} ] + \nonumber
\\
 && + i b_2 \left\{ \phantom{|}^{\mu\nu\alpha\beta}_{abcd} \right\}
[ \bar{\Psi}_{\mu\nu}{}^a \Gamma^{bcd}
\Psi_{\alpha\beta} -  \bar{\Psi}_{\mu\nu} \Gamma^{abc}
\Psi_{\alpha\beta}{}^d ] 
\end{eqnarray}
which is invariant under the following gauge transformations:
\begin{eqnarray}
\delta \Psi_{\mu\nu}{}^a &=& D_{[\mu} \xi_{\nu]}{}^a - 
\frac{i a_0}{5(d-4)} [ \gamma_{[\mu} \xi_{\nu]}{}^a + \frac{2}{d}
\gamma^a \xi_{[\mu,\nu]} ]  - \frac{b_2}{10(d-2)} [ e_{[\mu}{}^a
\xi_{\nu]} - \frac{1}{d} \gamma^a \gamma_{[\mu} \xi_{\nu]} ] \nonumber
\\
\delta \Psi_{\mu\nu} &=& D_{[\mu} \xi_{\nu]} + \frac{b_2}{20}
\xi_{[\mu,\nu]} + \frac{i a_2}{5(d-4)} \gamma_{[\mu} \xi_{\nu]}
\end{eqnarray}
At the same time, two other fields $\Phi_\mu{}^a$, $\Phi_\mu$
provides gauge invariant description for partially massless spin 5/2
particle \cite{Zin08c} with the Lagrangian:
\begin{eqnarray}
{\cal L} &=&  i \left\{ \phantom{|}^{\mu\nu\alpha}_{abc} \right\} [
\bar{\Phi}_\mu{}^d \Gamma^{abc} D_\nu \Phi_\alpha{}^d - 6
\bar{\Phi}_\mu{}^a \gamma^b D_\nu \Phi_\alpha{}^c -
\bar{\Phi}_\mu \Gamma^{abc} D_\nu \Phi_\alpha ] + \nonumber \\
 && + \left\{ \phantom{|}^{\mu\nu}_{ab} \right\} [
a_1 ( \bar{\Phi}_\mu{}^c \Gamma^{ab} \Phi_\nu{}^c + 2
\bar{\Phi}_\mu{}^a \Phi_\nu{}^b )  + a_3 \bar{\Phi}_\mu \Gamma^{ab}
\Phi_\nu ] + \nonumber \\
 && + i b_3 \left\{ \phantom{|}^{\mu\nu}_{ab} \right\} [
\bar{\Phi}_\mu{}^a \gamma^b \Phi_\nu -
\bar{\Phi}_\mu \gamma^a \Phi_\nu{}^b ]
\end{eqnarray}
which is invariant under the following gauge transformations:
\begin{eqnarray}
\delta \Phi_\mu{}^a &=& D_\mu \zeta^a + 
\frac{i a_1}{3(d-2)} [ \gamma_\mu \zeta^a - \frac{2}{d} \gamma^a
\zeta_\mu ] + \frac{b_3}{6(d-1)} [ e_\mu{}^a \zeta - \frac{1}{d}
\gamma^a \gamma_\mu \zeta ] \nonumber \\
\delta \Phi_\mu &=& D_\mu \zeta + \frac{b_3}{6} \zeta_\mu 
 - \frac{i a_3}{3(d-2)} \gamma_\mu \zeta 
\end{eqnarray}

\subsection{$Y(k+\frac{3}{2},\frac{3}{2})$}

We proceed with the construction of massive theory for spin-tensor
$\Psi_{\mu\nu}{}^{(a_k)}$ with arbitrary $k \ge 1$. Again our first
task to determine a set of fields necessary for gauge invariant
description of such massive field. Main gauge field 
$Y(k+\frac{3}{2},\frac{1}{2})$ has two gauge transformations with
parameters $Y(k+\frac{3}{2},\frac{1}{2})$ and
$Y(k+\frac{1}{2},\frac{3}{2})$ so we need two corresponding primary
fields. The first of them has one gauge transformation with parameter
$Y(k+\frac{1}{2},\frac{1}{2})$, while the second one has two gauge
transformations with parameters $Y(k-\frac{1}{2},\frac{3}{2})$ and
$Y(k+\frac{1}{2},\frac{1}{2})$. Taking into account reducibility of
gauge transformations of the main gauge field we have to introduce two
secondary fields $Y(k-\frac{1}{2},\frac{3}{2})$ and 
$Y(k+\frac{1}{2},\frac{1}{2})$ only. It is not hard to check that the
procedure again stops at the $Y(\frac{3}{2},\frac{1}{2})$ and
we need totally $Y(l+\frac{3}{2},\frac{3}{2})$ and 
$Y(l+\frac{3}{2},\frac{1}{2})$ with $0 \le l \le k$. Thus we introduce
the following fields: $\Psi_{\mu\nu}{}^{(a_l)}$ and
$\Phi_\mu{}^{(a_l)}$,  $0 \le l \le k$.

As we have already noted gauge invariant Lagrangian for massive
fermionic field contains kinetic and mass terms for all components as
well as cross terms without derivatives for all nearest neighbours.
Thus we will look for massive Lagrangian in the form:
$$
{\cal L} = \sum_{l=0}^k [ {\cal L}(\Psi_{\mu\nu}{}^{(a_l)}) + 
{\cal L}(\Phi_\mu{}^{(a_l)}) ] + \sum_{l=0}^{k-1} {\cal L}_{cross}(l)
$$
where
\begin{eqnarray}
(-1)^l {\cal L}(\Psi_{\mu\nu}{}^{(a_l)}) &=&
- i \left\{ \phantom{|}^{\mu\nu\alpha\beta\gamma}_{abcde}
\right\} [ \bar{\Psi}_{\mu\nu}{}^{(a_l)} \Gamma^{abcde} D_\alpha
\Psi_{\beta\gamma}{}^{(a_l)} - 10l \bar{\Psi}_{\mu\nu}{}^{a(a_{l-1})}
\Gamma^{bcd} D_\alpha \Psi_{\beta\gamma}{}^{e(a_{l-1})} ] + \nonumber
\\
 && + a_l \left\{ \phantom{|}^{\mu\nu\alpha\beta}_{abcd}
\right\} [ \bar{\Psi}_{\mu\nu}{}^{(a_l)} \Gamma^{abcd} 
\Psi_{\alpha\beta}{}^{(a_l)} + 6l \bar{\Psi}_{\mu\nu}{}^{a(a_{l-1})}
\Gamma^{bc} \Psi_{\alpha\beta}{}^{d(a_{l-1})} ]
\end{eqnarray}
\begin{eqnarray}
(-1)^l {\cal L}(\Phi_\mu{}^{(a_l)}) &=& - i \left\{
\phantom{|}^{\mu\nu\alpha}_{abc}
\right\} [ \bar{\Phi}_\mu{}^{(a_l)} \Gamma^{abc} D_\nu
\Phi_\alpha{}^{(a_l)} - 6l \bar{\Phi}_\mu{}^{a(a_{l-1})} \gamma^b
D_\nu \Phi_\alpha{}^{c(a_{l-1})} ] + \nonumber \\
 && + b_l \left\{ \phantom{|}^{\mu\nu}_{ab} \right\} [
\bar{\Phi}_\mu{}^{(a_l)} \Gamma^{ab} \Phi_\nu{}^{(a_l)} + 2l
\bar{\Phi}_\mu{}^{a(a_{l-1})} \Phi_\nu{}^{b(a_{l-1})} ]
\end{eqnarray}
\begin{eqnarray}
- i (-1)^l {\cal L}_{cross}(l) &=& 
c_l \left\{ \phantom{|}^{\mu\nu\alpha\beta}_{abcd} \right\} [
\bar{\Psi}_{\mu\nu}{}^{a(a_l)} \Gamma^{bcd} 
\Psi_{\alpha\beta}{}^{(a_l)} - \bar{\Psi}_{\mu\nu}{}^{(a_l)}
\Gamma^{abc} \Psi_{\alpha\beta}{}^{d(a_l)} ] + \nonumber \\
 && + d_l \left\{ \phantom{|}^{\mu\nu\alpha}_{abc} \right\} [
\bar{\Psi}_{\mu\nu}{}^{(a_l)} \Gamma^{abc} \Phi_\alpha{}^{(a_l)} -
6l \bar{\Psi}_{\mu\nu}{}^{a(a_{l-1})} \gamma^b 
\Phi_\alpha{}^{c(a_{l-1})} ] + \nonumber \\
 && + d_l \left\{ \phantom{|}^{\mu\nu\alpha}_{abc} \right\} [
\bar{\Phi}_\mu{}^{(a_l)} \Gamma^{abc} \Psi_{\nu\alpha}{}^{(a_l)} -
6l \Phi_\mu{}^{a(a_{l-1})} \gamma^b \Psi_{\nu\alpha}{}^{c(a_{l-1})} ]
+ \nonumber \\
 && + e_l \left\{ \phantom{|}^{\mu\nu}_{ab} \right\} [
\bar{\Phi}_\mu{}^{a(a_l)} \gamma^b \Phi_\nu{}^{(a_l)} -
\bar{\Phi}_\mu{}^{(a_l)} \gamma^a \Phi_\nu{}^{b(a_l)} ]
\end{eqnarray}

As usual, to compensate non-invariance of all mass terms (both
diagonal as well as cross terms) under the initial gauge
transformations, we have to introduce corresponding corrections to
gauge transformations. We have already introduced such corrections for
diagonal mass terms with coefficients $a_l$ and $b_l$ in Subsection
2.3 and Subsection 2.1 respectively. Let us consider three possible
type of cross terms in turn.

$\Psi_{\mu\nu}{}^{(a_{l+1})} \Leftrightarrow \Psi_{\mu\nu}{}^{(a_l)}$.
In this case cross terms look as:
$$
\Delta {\cal L} = - i (-1)^l c_l \left\{
\phantom{|}^{\mu\nu\alpha\beta}_{abcd} \right\} [
\bar{\Psi}_{\mu\nu}{}^{a(a_l)} \Gamma^{bcd} 
\Psi_{\alpha\beta}{}^{(a_l)} - \bar{\Psi}_{\mu\nu}{}^{(a_l)}
\Gamma^{abc} \Psi_{\alpha\beta}{}^{d(a_l)} ] 
$$
and to compensate for their non-invariance we have to introduce:
\begin{eqnarray}
\delta' \Psi_{\mu\nu}{}^{(a_{l+1})} &=& - \frac{c_l}{10(l+1)(d+l-2)} [
e_{[\mu}{}^{(a_1} \xi_{\nu]}{}^{a_l)} - \frac{1}{(d+2l)} \gamma^{(a_1}
\gamma_{[\mu} \xi_{\nu]}{}^{a_l)} + \nonumber \\
 && \qquad \qquad \qquad \qquad \qquad +  \frac{2}{(d+2l)} g^{(a_1a_2}
\xi_{[\mu,\nu]}{}^{a_{l-1})} ] \nonumber \\
\delta' \Psi_{\mu\nu}{}^{(a_l)} &=& \frac{c_l}{10(l+2)} 
\xi_{[\mu,\nu]}{}^{(a_l)}
\end{eqnarray}

$\Psi_{\mu\nu}{}^{(a_l)} \Leftrightarrow \Phi_\mu{}^{(a_l)}$.
Here the cross terms have the following form:
$$
\Delta {\cal L} = - i (-1)^l d_l \left\{
\phantom{|}^{\mu\nu\alpha}_{abc} \right\} [
\bar{\Psi}_{\mu\nu}{}^{(a_l)} \Gamma^{abc} \Phi_\alpha{}^{(a_l)} -
6l \bar{\Psi}_{\mu\nu}{}^{a(a_{l-1})} \gamma^b 
\Phi_\alpha{}^{c(a_{l-1})} ] + h.c.
$$
and to compensate for their non-invariance we have to introduce the
following corrections:
\begin{eqnarray}
\delta' \Psi_{\mu\nu}{}^{(a_l)} &=& - 
\frac{(l+1) d_l}{10(l+2)(d-3)(d-4)} [ \Gamma_{\mu\nu} \zeta^{(a_l)} -
\frac{(d-3)(d-4) + 2l +2}{2(l+1)(d+l-3)} e_{[\mu}{}^{(a_1} 
\zeta_{\mu]}{}^{a_{l-1})} +  \nonumber \\
 && \qquad \qquad \qquad \qquad \qquad + 
\frac{(d+2l-1)}{2(l+1)(d+l-3)} \gamma^{(a_1}
\gamma_{[\mu} \zeta_{\nu]}{}^{a_{l-1})} ] \nonumber \\
\delta' \Phi_\mu{}^{(a_l)} &=& 2 d_l \xi_\mu{}^{(a_l)}
\end{eqnarray}

$\Phi_\mu{}^{(a_{l+1})} \Leftrightarrow \Phi_\mu{}^{(a_l)}$.
The last possible type of cross terms have the form:
$$
\Delta {\cal L} = - i (-1)^l e_l \left\{ \phantom{|}^{\mu\nu}_{ab}
\right\} [ \bar{\Phi}_\mu{}^{a(a_l)} \gamma^b \Phi_\nu{}^{(a_l)} -
\bar{\Phi}_\mu{}^{(a_l)} \gamma^a \Phi_\nu{}^{b(a_l)} ]
$$
while corrections to gauge transformations can be written as follows:
\begin{eqnarray}
\delta' \Phi_\mu{}^{(a_{l+1})} &=&  \frac{e_l}{6(l+1)(d+l-1)} [
e_\mu{}^{(a_1} \zeta^{a_l)} - \frac{1}{(d+2l)} \gamma^{(a_1}
\gamma_\mu \zeta^{a_l)} - \frac{2}{(d+2l)} g^{(a_1 a_2}
\zeta_\mu{}^{a_{l-1})} ]  \nonumber \\
\delta' \Phi_\mu{}^{(a_l)} &=& \frac{e_l}{6(l+1)} \zeta^{(a_l)}
\end{eqnarray}

Collecting all pieces together we obtain the following complete set of
gauge transformations (for simplicity we omit here complicated terms
which are necessary to ensure that all variations are
$\gamma$-transverse):
\begin{eqnarray}
\delta \Psi_{\mu\nu}{}^{(a_l)} &=& D_{[\mu} \xi_{\nu]}{}^{(a_l)} +
\frac{i a_l}{5(d-4)} [ \gamma_{[\mu} \xi_{\nu]}{}^{(a_l)} + \dots] +
\frac{c_l}{10(l+2)} \xi_{[\mu,\nu]}{}^{(a_l)} - \nonumber \\
 && - \frac{c_{l-1}}{10l(d+l-3)} [ e_{[\mu}{}^{(a_1} 
\xi_{\nu]}{}^{a_{l-1})} + \dots] - \frac{(l+1) d_l}{10(l+2)(d-3)(d-5)}
[ \Gamma_{\mu\nu} \zeta^{(a_l)} + \dots] \nonumber \\
\delta \Phi_\mu{}^{(a_l)} &=& D_\mu \zeta^{(a_l)} - 
\frac{i b_l}{3(d-2)} [ \gamma_\mu \zeta^{(a_l)} + \dots] + 2 d_l
\xi_\mu{}^{(a_l)} + \frac{e_l}{6(l+1)} \zeta^{(a_l)} + \\
 && + \frac{e_{l-1}}{6l(d+l-2)} [ e_\mu{}^{(a_1} \zeta^{a_{l-1})} +
\dots ] \nonumber
\end{eqnarray}

At this stage we have complete Lagrangian and gauge transformations,
in this all parameters in gauge transformations are expressed in terms
of the Lagrangian ones $a$, $b$, $c$, $d$ and $e$ so that all
variations with one derivative cancel. Our next task --- calculate all
variations without derivatives (including contribution of kinetic
terms due to non-commutativity of covariant derivatives) and require
their cancellation. We will not give here these lengthy but
straightforward calculations presenting final results only. First of
all we obtain a number of recurrent relations on diagonal mass
parameters $a_l$ and $b_l$ which allows us to express all of them in
terms of the main one $a_k = M$:
$$
a_l = \frac{(d+2k)}{(d+2l)} M, \qquad
b_l = - \frac{3(d-2)}{5(d-4)} a_l
$$
Then we obtain recurrent relations on the parameters $d_l$ which
allows us to express all of them in terms of main one $d_k = m$ (it is
not a mass, just notation):
$$
d_l{}^2 = \frac{(k+1)(d+k-2)}{(l+1)(d+l-2)} m^2
$$
Further we get the following expressions for the parameters $c_l$ and
$e_l$:
$$
c_l{}^2 = \frac{10(k-l)(l+1)(d+k+l)}{(d+2l)} [
\frac{(k+1)(d-4)}{(k+2)(d-3)} m^2 + 10 (l+2) (d+l-2) \kappa ]
$$
$$
e_l{}^2 = \frac{9(l+1)(d+l-1)}{25(l+2)(d+l-2)} c_l{}^2
$$
At last we obtain an important relation on parameters $M$ and $m$:
$$
M^2 = \frac{5(k+1)(d-4)}{2(k+2)(d-3)} m^2 - 
\frac{25}{4} (d-4)^2 \kappa
$$

Now we are ready to analyze the results obtained. To clarify the roles
played by parameters $c$, $d$ and $e$ we give here Figure 4.
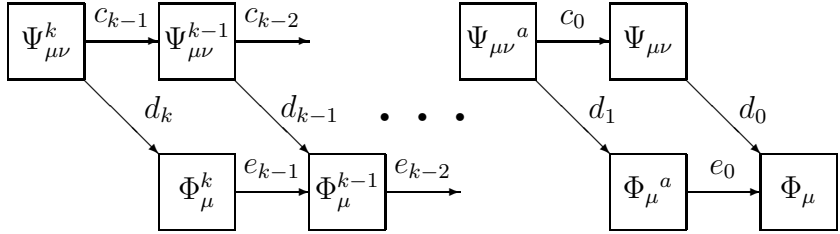
\begin{figure}[htb]
\begin{center}
\begin{picture}(140,32)
\put(10,21){\framebox(10,10)[]{$\Psi_{\mu\nu}^{k}$}}
\put(30,21){\framebox(10,10)[]{$\Psi_{\mu\nu}^{k-1}$}}
\put(30,1){\framebox(10,10)[]{$\Phi_\mu^{k}$}}
\put(50,1){\framebox(10,10)[]{$\Phi_\mu^{k-1}$}}
\put(20,26){\vector(1,0){10}}
\put(22,26){\makebox(6,6)[]{$c_{k-1}$}}
\put(40,26){\vector(1,0){10}}
\put(42,26){\makebox(6,6)[]{$c_{k-2}$}}
\put(20,21){\vector(1,-1){10}}
\put(27,14){\makebox(6,6)[]{$d_k$}}
\put(40,21){\vector(1,-1){10}}
\put(47,14){\makebox(6,6)[]{$d_{k-1}$}}
\put(40,6){\vector(1,0){10}}
\put(42,6){\makebox(6,6)[]{$e_{k-1}$}}
\put(60,6){\vector(1,0){10}}
\put(62,6){\makebox(6,6)[]{$e_{k-2}$}}
\multiput(60,16)(5,0){3}{\circle*{1}}
\put(70,21){\framebox(10,10)[]{$\Psi_{\mu\nu}{}^a$}}
\put(90,21){\framebox(10,10)[]{$\Psi_{\mu\nu}$}}
\put(90,1){\framebox(10,10)[]{$\Phi_\mu{}^a$}}
\put(110,1){\framebox(10,10)[]{$\Phi_\mu$}}
\put(80,26){\vector(1,0){10}}
\put(82,26){\makebox(6,6)[]{$c_0$}}
\put(80,21){\vector(1,-1){10}}
\put(86,14){\makebox(6,6)[]{$d_1$}}
\put(100,21){\vector(1,-1){10}}
\put(106,14){\makebox(6,6)[]{$d_0$}}
\put(100,6){\vector(1,0){10}}
\put(102,6){\makebox(6,6)[]{$e_0$}}
\end{picture}
\end{center}
\caption{General massive $Y(k+\frac{3}{2},\frac{3}{2})$ theory}
\end{figure}
First of all note that to obtain massless limit we have to put $m \to
0$ and $c_{k-1} \to 0$ simultaneously. But as the expression on
$c_{k-1}$ clearly shows such limit is possible in flat Minkowski space
($\kappa = 0$) only. For non-zero values of $\kappa$ we can obtain a
number of partially massless limits. Let us consider $AdS$ space
($\kappa < 0$) first. The most physically interesting limit appears
then $c_{k-1} \to 0$ (and hence $e_{k-1} \to 0$). In this the whole
system decomposes into two disconnected subsystems as shown on the
Figure 5.
\begin{figure}[htb]
\begin{center}
\begin{picture}(140,32)
\put(10,21){\framebox(10,10)[]{$\Psi_{\mu\nu}^{k}$}}
\put(30,21){\framebox(10,10)[]{$\Psi_{\mu\nu}^{k-1}$}}
\put(30,1){\framebox(10,10)[]{$\Phi_\mu^{k}$}}
\put(50,1){\framebox(10,10)[]{$\Phi_\mu^{k-1}$}}
\put(40,26){\vector(1,0){10}}
\put(42,26){\makebox(6,6)[]{$c_{k-2}$}}
\put(20,21){\vector(1,-1){10}}
\put(27,14){\makebox(6,6)[]{$d_k$}}
\put(40,21){\vector(1,-1){10}}
\put(47,14){\makebox(6,6)[]{$d_{k-1}$}}
\put(60,6){\vector(1,0){10}}
\put(62,6){\makebox(6,6)[]{$e_{k-2}$}}
\multiput(60,16)(5,0){3}{\circle*{1}}
\put(70,21){\framebox(10,10)[]{$\Psi_{\mu\nu}{}^a$}}
\put(90,21){\framebox(10,10)[]{$\Psi_{\mu\nu}$}}
\put(90,1){\framebox(10,10)[]{$\Phi_\mu{}^a$}}
\put(110,1){\framebox(10,10)[]{$\Phi_\mu$}}
\put(80,26){\vector(1,0){10}}
\put(82,26){\makebox(6,6)[]{$c_0$}}
\put(80,21){\vector(1,-1){10}}
\put(86,14){\makebox(6,6)[]{$d_1$}}
\put(100,21){\vector(1,-1){10}}
\put(106,14){\makebox(6,6)[]{$d_0$}}
\put(100,6){\vector(1,0){10}}
\put(102,6){\makebox(6,6)[]{$e_0$}}
\end{picture}
\end{center}
\caption{Unitary partially massless limit in $AdS$ space}
\end{figure}
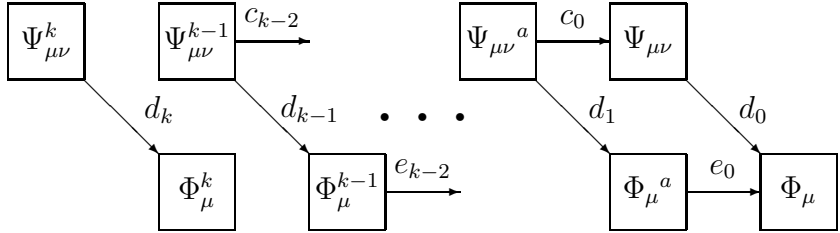
In this, two fields $\Psi_{\mu\nu}{}^{(a_k)}$ and $\Phi_\mu{}^{(a_k)}$
describe partially massless theory corresponding to unitary
irreducible representation of $AdS$ group \cite{BMV00}. The Lagrangian
for such theory has the form:
\begin{eqnarray}
(-1)^k {\cal L} &=& - i \left\{
\phantom{|}^{\mu\nu\alpha\beta\gamma}_{abcde}
\right\} [ \bar{\Psi}_{\mu\nu}{}^{(a_k)} \Gamma^{abcde} D_\alpha
\Psi_{\beta\gamma}{}^{(a_k)} - 10k \bar{\Psi}_{\mu\nu}{}^{a(a_{k-1})}
\Gamma^{bcd} D_\alpha \Psi_{\beta\gamma}{}^{e(a_{k-1})} ] + \nonumber
\\
&&  - i \left\{
\phantom{|}^{\mu\nu\alpha}_{abc}
\right\} [ \bar{\Phi}_\mu{}^{(a_k)} \Gamma^{abc} D_\nu
\Phi_\alpha{}^{(a_k)} - 6k \bar{\Phi}_\mu{}^{a(a_{k-1})} \gamma^b
D_\nu \Phi_\alpha{}^{c(a_{k-1})} ] + \nonumber \\
 && + a_k \left\{ \phantom{|}^{\mu\nu\alpha\beta}_{abcd}
\right\} [ \bar{\Psi}_{\mu\nu}{}^{(a_k)} \Gamma^{abcd} 
\Psi_{\alpha\beta}{}^{(a_k)} + 6k \bar{\Psi}_{\mu\nu}{}^{a(a_{k-1})}
\Gamma^{bc} \Psi_{\alpha\beta}{}^{d(a_{k-1})} ] + \nonumber \\
 && + b_k \left\{ \phantom{|}^{\mu\nu}_{ab} \right\} [
\bar{\Phi}_\mu{}^{(a_k)} \Gamma^{ab} \Phi_\nu{}^{(a_k)} + 2k
\bar{\Phi}_\mu{}^{a(a_{k-1})} \Phi_\nu{}^{b(a_{k-1})} ] + \nonumber \\
 && + d_k \left\{ \phantom{|}^{\mu\nu\alpha}_{abc} \right\} [
\bar{\Psi}_{\mu\nu}{}^{(a_k)} \Gamma^{abc} \Phi_\alpha{}^{(a_k)} -
6k \bar{\Psi}_{\mu\nu}{}^{a(a_{k-1})} \gamma^b 
\Phi_\alpha{}^{c(a_{k-1})} ] + \nonumber \\
 && + d_k \left\{ \phantom{|}^{\mu\nu\alpha}_{abc} \right\} [
\bar{\Phi}_\mu{}^{(a_k} \Gamma^{abc} \Psi_{\nu\alpha}{}^{(a_k)} -
6k \Phi_\mu{}^{a(a_{k-1})} \gamma^b \Psi_{\nu\alpha}{}^{c(a_{k-1})} ]
\end{eqnarray}
while gauge transformations leaving it invariant look as follows:
\begin{eqnarray}
\delta \Psi_{\mu\nu}{}^{(a_k)} &=& D_{[\mu} \xi_{\nu]}{}^{(a_k)} +
\frac{i a_k}{5(d-4)} [ \gamma_{[\mu} \xi_{\nu]}{}^{(a_k)} + \dots]
 - \nonumber \\
 && - \frac{(l+1) d_k}{10(k+2)(d-3)(d-5)}
[ \Gamma_{\mu\nu} \zeta^{(a_k)} + \dots]  \\
\delta \Phi_\mu{}^{(a_k)} &=& D_\mu \zeta^{(a_k)} - 
\frac{i b_k}{3(d-2)} [ \gamma_\mu \zeta^{(a_k)} + \dots] + 2 d_k
\xi_\mu{}^{(a_k)} \nonumber
\end{eqnarray}
At the same time all other fields just give massive theory for the
$\Psi_{\mu\nu}{}^{(a_{k-1})}$ spin-tensor. Besides a number of
non-unitary partially massless limits exists. Indeed, each time when
one of the $c_l \to 0$ (and hence $e_l \to 0$) the whole system also
decomposes into two disconnected subsystems. One of them with the
fields $\Psi_{\mu\nu}{}^{(a_m)}$ and $\Phi_\mu{}^{(a_m)}$ with $l \le
m \le k$ describes a non-unitary partially massless theory, while
remaining fields just give massive theory for the 
$\Psi_{\mu\nu}{}^{(a_{l-1})}$ spin-tensor.

Let us turn to the $dS$ space ($\kappa > 0$). From the last relation
on parameters $M$ and $m$ we see that there is a unitary forbidden
region $m^2 < \frac{5(k+2)(d-3)(d-4)}{2(k+1)} \kappa$. Inside this
region lives the only partially massless limit possible. It appears
then we put $m \to 0$ (and this puts all $d_l \to 0$ simultaneously).
Once again the whole system decomposes into two disconnected parts as
shown on the Figure 6.
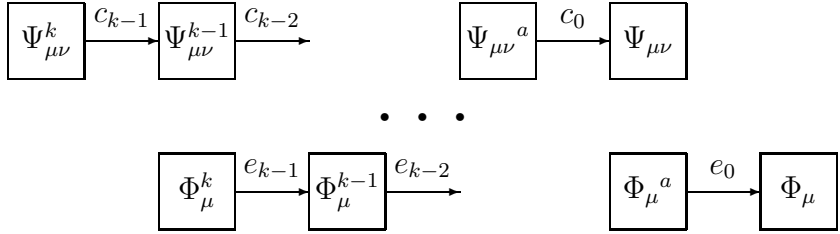
\begin{figure}[htb]
\begin{center}
\begin{picture}(140,32)
\put(10,21){\framebox(10,10)[]{$\Psi_{\mu\nu}^{k}$}}
\put(30,21){\framebox(10,10)[]{$\Psi_{\mu\nu}^{k-1}$}}
\put(30,1){\framebox(10,10)[]{$\Phi_\mu^{k}$}}
\put(50,1){\framebox(10,10)[]{$\Phi_\mu^{k-1}$}}
\put(20,26){\vector(1,0){10}}
\put(22,26){\makebox(6,6)[]{$c_{k-1}$}}
\put(40,26){\vector(1,0){10}}
\put(42,26){\makebox(6,6)[]{$c_{k-2}$}}
\put(40,6){\vector(1,0){10}}
\put(42,6){\makebox(6,6)[]{$e_{k-1}$}}
\put(60,6){\vector(1,0){10}}
\put(62,6){\makebox(6,6)[]{$e_{k-2}$}}
\multiput(60,16)(5,0){3}{\circle*{1}}
\put(70,21){\framebox(10,10)[]{$\Psi_{\mu\nu}{}^a$}}
\put(90,21){\framebox(10,10)[]{$\Psi_{\mu\nu}$}}
\put(90,1){\framebox(10,10)[]{$\Phi_\mu{}^a$}}
\put(110,1){\framebox(10,10)[]{$\Phi_\mu$}}
\put(80,26){\vector(1,0){10}}
\put(82,26){\makebox(6,6)[]{$c_0$}}
\put(100,6){\vector(1,0){10}}
\put(102,6){\makebox(6,6)[]{$e_0$}}
\end{picture}
\end{center}
\caption{Non-unitary partially massless limit in $dS$ space}
\end{figure}
One of them with the fields $\Psi_{\mu\nu}{}^{(a_l)}$ $0 \le l \le k$
provides one more example of partially massless theory in $dS$ space
with the Lagrangian
\begin{eqnarray}
{\cal L} &=& \sum_{l=0}^k (-1)^l \left[
- i \left\{ \phantom{|}^{\mu\nu\alpha\beta\gamma}_{abcde}
\right\} [ \bar{\Psi}_{\mu\nu}{}^{(a_l)} \Gamma^{abcde} D_\alpha
\Psi_{\beta\gamma}{}^{(a_l)} - 10l \bar{\Psi}_{\mu\nu}{}^{a(a_{l-1})}
\Gamma^{bcd} D_\alpha \Psi_{\beta\gamma}{}^{e(a_{l-1})} ] + \right.
\nonumber \\
 && \qquad \qquad \left. + a_l \left\{
\phantom{|}^{\mu\nu\alpha\beta}_{abcd}
\right\} [ \bar{\Psi}_{\mu\nu}{}^{(a_l)} \Gamma^{abcd} 
\Psi_{\alpha\beta}{}^{(a_l)} + 6l \bar{\Psi}_{\mu\nu}{}^{a(a_{l-1})}
\Gamma^{bc} \Psi_{\alpha\beta}{}^{d(a_{l-1})} ] \right] + \nonumber \\
 && + i \sum_{l=0}^{k-1} (-1)^l c_l \left\{
\phantom{|}^{\mu\nu\alpha\beta}_{abcd}
\right\} [ \bar{\Psi}_{\mu\nu}{}^{a(a_l)} \Gamma^{bcd} 
\Psi_{\alpha\beta}{}^{(a_l)} - \bar{\Psi}_{\mu\nu}{}^{(a_l)}
\Gamma^{abc} \Psi_{\alpha\beta}{}^{d(a_l)} ]
\end{eqnarray}
which is invariant under the following gauge transformations:
\begin{eqnarray}
\delta \Psi_{\mu\nu}{}^{(a_l)} &=& D_{[\mu} \xi_{\nu]}{}^{(a_l)} +
\frac{i a_l}{5(d-4)} [ \gamma_{[\mu} \xi_{\nu]}{}^{(a_l)} + \dots] +
\frac{c_l}{10(l+2)} \xi_{[\mu,\nu]}{}^{(a_l)} - \nonumber \\
 && - \frac{c_{l-1}}{10l(d+l-3)} [ e_{[\mu}{}^{(a_1} 
\xi_{\nu]}{}^{a_{l-1})} + \dots]
\end{eqnarray}
In this, remaining fields $\Phi_\mu{}^{(a_l)}$ $0 \le l \le k$ realize
partially massless theory constructed earlier \cite{Zin08b}.

\subsection{$Y(\frac{5}{2},\frac{5}{2})$}

As we have already noted, (spin)-tensors corresponding to Young
tableau
with equal number of boxes in both rows are special and require
separate consideration. Here we consider simplest example --- massive
theory for $R_{\mu\nu}{}^{ab}$ spin-tensor. First of all we have to
find a set of fields necessary for gauge invariant description of such
massive spin-tensor. Our main gauge field $Y(\frac{5}{2},\frac{5}{2})$
has one own gauge transformation with the parameter 
$Y(\frac{5}{2},\frac{3}{2})$ only (and this is a main feature making
this (spin)-tensors special). Thus we need one primary Goldstone field
$Y(\frac{5}{2},\frac{3}{2})$ only. This field has two gauge
transformations with parameters $Y(\frac{5}{2},\frac{1}{2})$ and
$Y(\frac{3}{2},\frac{3}{2})$ but due to reducibility of our main field
gauge transformations we need one secondary field 
$Y(\frac{5}{2},\frac{1}{2})$ only. This field also has one own gauge
transformation with the parameter $Y(\frac{3}{2},\frac{1}{2})$ but due
to reducibility of primary field gauge transformations the procedure
stops here. Thus we need three fields $R_{\mu\nu}{}^{ab}$,
$\Psi_{\mu\nu}{}^a$ and $\Phi_\mu{}^a$ only.

As in all previous cases, we will construct a massive gauge invariant
Lagrangian as the sum of kinetic and mass terms for all three fields
as well as cross terms without derivatives:
\begin{eqnarray}
{\cal L} &=& - i \left\{ \phantom{|}^{\mu\nu\alpha\beta\gamma}_{abcde}
\right\} [ \bar{R}_{\mu\nu}{}^{fg} \Gamma^{abcde} D_\alpha
R_{\beta\gamma}{}^{fg} - 20 \bar{R}_{\mu\nu}{}^{af} \Gamma^{bcd}
D_\alpha R_{\beta\gamma}{}^{ef} - 60 \bar{R}_{\mu\nu}{}^{ab} \gamma^c
D_\alpha R_{\beta\gamma}{}^{de} ] + \nonumber \\
 && + i \left\{ \phantom{|}^{\mu\nu\alpha\beta\gamma}_{abcde}
\right\} [ \bar{\Psi}_{\mu\nu}{}^f \Gamma^{abcde} D_\alpha
\Psi_{\beta\gamma}{}^f - 10 \bar{\Psi}_{\mu\nu}{}^a \Gamma^{bcd}
D_\alpha \Psi_{\beta\gamma}{}^e ] + \nonumber \\
 && + i \left\{ \phantom{|}^{\mu\nu\alpha}_{abc} \right\} [
\bar{\Phi}_\mu{}^d \Gamma^{abc} D_\nu \Phi_\alpha{}^d - 6
\bar{\Phi}_\mu{}^a \gamma^b D_\nu \Phi_\alpha{}^c ] + \nonumber \\ 
&& + a_0 \left\{ \phantom{|}^{\mu\nu\alpha\beta}_{abcd} \right\} [
 \bar{R}_{\mu\nu}{}^{ef} \Gamma^{abcd} R_{\alpha\beta}{}^{ef} + 12
\bar{R}_{\mu\nu}{}^{ae} \Gamma^{bc} R_{\alpha\beta}{}^{de} - 12
\bar{R}_{\mu\nu}{}^{ab} R_{\alpha\beta}{}^{cd} ] + \nonumber \\
 && + a_1 \left\{ \phantom{|}^{\mu\nu\alpha\beta}_{abcd} \right\} [
\bar{\Psi}_{\mu\nu}{}^e \Gamma^{abcd} \Psi_{\alpha\beta}{}^e + 6
\bar{\Psi}_{\mu\nu}{}^a \Gamma^{bc} \Psi_{\alpha\beta}{}^d ] +
a_2 \left\{ \phantom{|}^{\mu\nu}_{ab} \right\} [
 \bar{\Phi}_\mu{}^c \Gamma^{ab} \Phi_\nu{}^c + 2
\bar{\Phi}_\mu{}^a \Phi_\nu{}^b ] + \nonumber \\
 && + i b_1 \left\{ \phantom{|}^{\mu\nu\alpha\beta}_{abcd} \right\} [
\bar{R}_{\mu\nu}{}^{ae} \Gamma^{bcd} \Psi_{\alpha\beta}{}^e - 6
\bar{R}_{\mu\nu}{}^{ab} \gamma^c \Psi_{\alpha\beta}{}^d -
\bar{\Psi}_{\mu\nu}{}^e \Gamma^{abc} R_{\alpha\beta}{}^{de} - 6
\bar{\Psi}_{\mu\nu}{}^a \gamma^b R_{\alpha\beta}{}^{cd} ] + \nonumber
\\
 && + i b_2 \left\{ \phantom{|}^{\mu\nu\alpha}_{abc} \right\} [
\bar{\Psi}_{\mu\nu}{}^d \Gamma^{abc} \Phi_\alpha{}^d - 6
\bar{\Psi}_{\mu\nu}{}^a \gamma^b \Phi_\alpha{}^c + 
\bar{\Phi}_\mu{}^d \Gamma^{abc} \Psi_{\nu\alpha}{}^d - 6
\bar{\Phi}_\mu{}^a \gamma^b \Psi_{\nu\alpha}{}^c ]
\end{eqnarray}
Now following our usual strategy we calculate variations with one
derivative to find appropriate corrections to gauge transformations.
Really most of them we are already familiar with, the only new ones
are related with the cross terms $R_{\mu\nu}{}^{ab} \Leftrightarrow
\Psi_{\mu\nu}{}^a$. Calculating these new corrections and collecting
previously known results we obtain:
\begin{eqnarray}
\delta R_{\mu\nu}{}^{ab} &=& D_{[\mu} \xi_{\nu]}{}^{ab} + 
\frac{i a_0}{5(d-4)} [ \gamma_{[\mu} \xi_{\nu]}{}^{ab} +
\frac{2}{(d-2)} \gamma^{[a} \xi_{[\mu,\nu]}{}^{b]} ] + \nonumber \\
 && + \frac{b_1}{20(d-3)} [ e_{[\mu}{}^{[a} \xi_{\nu]}{}^{b]}
- \frac{1}{(d-2)} \gamma^{[a} \gamma_{[\mu} \xi_{\nu]}{}^{b]} -
\frac{2}{(d-1)(d-2)} \Gamma^{ab} \xi_{[\mu,\nu]} ] \nonumber \\
\delta \Psi_{\mu\nu}{}^a &=& D_{[\mu} \xi_{\nu]}{}^a - \frac{b_1}{10}
\xi_{[\mu,\nu]}{}^a - \frac{i a_1}{5(d-4)} [ \gamma_{[\mu} 
\xi_{\nu]}{}^a + \frac{2}{d} \gamma^a \xi_{[\mu,\nu]} ] + \\
 && +  \frac{b_2}{15(d-3)(d-4)} [ \Gamma_{\mu\nu}
\zeta^a - \frac{(d^2-7d+16)}{4(d-2)} e_{[\mu}{}^a \zeta_{\nu]} +
\frac{(d+1)}{4(d-2)} \gamma^a
\gamma_{[\mu} \zeta_{\nu]} ] \nonumber \\
\delta \Phi_\mu{}^a &=& D_\mu \zeta^a - 2 b_2 \xi_\mu{}^a + 
\frac{a_2}{3(d-2)} [ \gamma_\mu \zeta^a - \frac{2}{d} \gamma^a
\zeta_\mu ] \nonumber
\end{eqnarray}
Now we proceed with the variations without derivatives (including
contribution of kinetic terms due to non-commutativity of covariant
derivatives). First of all, their cancellation leads to the relations
on the diagonal mass terms:
$$
a_1 = - \frac{d}{(d-2)} a_0, \qquad
a_2 = - \frac{3(d-2)}{5(d-4)} a_1 = \frac{3d}{5(d-4)} a_0
$$
Also we obtain two important relations:
$$
a_0{}^2 = \frac{(d-2)}{8(d-1)} b_1{}^2 - \frac{25(d-2)^2}{4} \kappa
$$
$$
b_2{}^2 = \frac{3(d-2)}{20(d-1)(d-4)} [ (d-2) b_1{}^2 - 200 (d-1)(d-3)
\kappa ]
$$
Simple linear structure of this theory $R_{\mu\nu}{}^{ab}
\Leftrightarrow \Psi_{\mu\nu}{}^a \Leftrightarrow \Phi_\mu{}^a$ makes
an analysis also simple. First of all we see that massless limit (i.e.
decoupling of $\Psi_{\mu\nu}{}^a$ from $R_{\mu\nu}{}^{ab}$)
corresponds to $b_1 \to 0$. Such a limit is possible in the $AdS$
space $\kappa < 0$ (and in the flat Minkowski space, of course) in
complete agreement with the fact that massless theory for 
$R_{\mu\nu}{}^{ab}$ admits deformation into $AdS$ space without
introduction of any additional fields. In this, two other fields
$\Psi_{\mu\nu}{}^a$ and $\Phi_\mu{}^a$ describe partially massless
theory we already familiar with. In the $dS$ space we once again face
an unitary forbidden region $b_1{}^2 < \frac{25}{2} (d-1)(d-2)
\kappa$. Inside this region we find one more example of non-unitary
partially massless theory. It appears then $b_2 \to 0$, in this the
field $\Phi_\mu{}^a$ decouples, while two other fields
$R_{\mu\nu}{}^{ab}$ and $\Psi_{\mu\nu}{}^a$ describe partially
massless theory. The Lagrangian and gauge transformations for this
theory can be easily obtained from the general formulas simply
omitting the field $\Phi_\mu{}^a$ and all terms in the gauge
transformations containing $\zeta^a$.

\subsection{$Y(k+\frac{3}{2},k+\frac{3}{2})$}

Let us consider now  general case --- spin-tensor 
$Y(k+\frac{3}{2},k+\frac{3}{2})$ with arbitrary $k \ge 1$. Again it is
crucial that the main field $R_{\mu\nu}{}^{(a_k),(b_k)}$ has one gauge
transformation with parameter $Y(k+\frac{3}{2},k+\frac{1}{2})$ only so
we need one primary field. This field has two gauge transformations
with parameters $Y(k+\frac{3}{2},k-\frac{1}{2})$ and 
$Y(k+\frac{1}{2},k+\frac{1}{2})$ but due to reducibility of gauge
transformations of main field we need one secondary field
$Y(k+\frac{3}{2},k+\frac{1}{2})$ only. It is not hard to check that
complete set of fields necessary for gauge invariant description
contains $Y(k+\frac{3}{2},l+\frac{3}{2})$ $0 \le l \le k$ and
$Y(k+\frac{3}{2},\frac{1}{2})$. 

Following our general procedure we will look for massive gauge
invariant Lagrangian as the sum of kinetic and mass terms for all
fields as well as cross terms for nearest neighbours:
\begin{equation}
{\cal L} = {\cal L}(R_{\mu\nu}{}^{(a_k),(b_k)}) + 
\sum_{l=0}^{k-1} {\cal L}(\Psi_{\mu\nu}{}^{(a_k),(b_l)}) +
{\cal L}(\Phi_\mu{}^{(a_k)}) + {\cal L}_{cross}
\end{equation}
where Lagrangian ${\cal L}(R_{\mu\nu}{}^{(a)k),(b_k)})$ is given by
formulas (\ref{lagkk0}) and (\ref{lagkk1}) of Subsection 2.5, while
Lagrangian ${\cal L}(\Psi_{\mu\nu}{}^{(a_k),(b_l)})$ is given by
formulas (\ref{lagkl0}) and (\ref{lagkl1}) of Subsection 2.6. Here
\begin{eqnarray}
{\cal L}_{cross} &=& i d_{k,k} \left\{
\phantom{|}^{\mu\nu\alpha\beta}_{abcd}
\right\} [ \bar{R}_{\mu\nu}{}^{(a_k),a(b_{k-1})} \Gamma^{bcd}
\Psi_{\alpha\beta}{}^{(a_k),(b_{k-1})} + 6k
\bar{R}_{\mu\nu}{}^{a(a_{k-1}),b(b_{k-1})} \gamma^c 
\Psi_{\alpha\beta}{}^{d(a_{k-1}),(b_{k-1})} \nonumber \\
 && \qquad \qquad  - \bar{\Psi}_{\mu\nu}{}^{(a_k),(a_{k-1})}
\Gamma^{abc} R_{\alpha\beta}{}^{(a_k),d(b_{k-1})} + 6k
\bar{\Psi}_{\mu\nu}{}^{a(a_{k-1}),(b_{k-1})} \gamma^b
R_{\alpha\beta}{}^{c(a_{k-1}),d(b_{k-1})} ] + \nonumber \\
 && + \sum_{l=1}^{k-1} i (-1)^{k+l} d_{k,l} 
\left\{ \phantom{|}^{\mu\nu\alpha\beta}_{abcd} \right\} [
\bar{\Psi}_{\mu\nu}{}^{(a_k),a(b_{l-1})} \Gamma^{bcd}
\Psi_{\alpha\beta}{}^{(a_k),(b_{l-1})} + \nonumber \\
 && \qquad \qquad + 6k
\bar{\Psi}_{\mu\nu}{}^{a(a_{k-1}),b(b_{l-1})} \gamma^c 
\Psi_{\alpha\beta}{}^{d(a_{k-1}),(b_{l-1})}  -
\bar{\Psi}_{\mu\nu}{}^{(a_k),(b_{l-1})} \Gamma^{abc}
\Psi_{\alpha\beta}{}^{(a_k),d(b_{l-1})} + \nonumber \\
 && \qquad \qquad \qquad + 6k
\bar{\Psi}_{\mu\nu}{}^{a(a_{k-1}),(b_{l-1})} \gamma^b 
\Psi_{\alpha\beta}{}^{c(a_{k-1}),d(b_{l-1})} ] + \nonumber \\
&& + i (-1)^k d_{k,0} \left\{ \phantom{|}^{\mu\nu\alpha}_{abc}
\right\} [ \bar{\Psi}_{\mu\nu}{}^{(k)} \Gamma^{abc}
\Phi_\alpha{}^{(k)} - 6k  \bar{\Psi}_{\mu\nu}{}^{a(k-1)} \gamma^b
\Phi_\alpha{}^{c(k-1)} + \nonumber \\ 
&&  \qquad \qquad \qquad \qquad + \bar{\Phi}_\mu{}^{(a_k)}
\Gamma^{abc} \Psi_{\nu\alpha}{}^{(a_k)} - 6k
\bar{\Phi}_\mu{}^{a(a_{k-1})} \gamma^b 
\Psi_{\nu\alpha}{}^{c(a_{k-1})} ]
\end{eqnarray}
As usual, to compensate for non-invariance of cross terms under the
initial gauge transformations, we have to introduce corresponding
corrections to gauge transformations. Let us consider different cross
terms in turn.

$R_{\mu\nu}{}^{(a_k),(b_k)} \Leftrightarrow 
\Psi_{\mu\nu}{}^{(a_k),(b_{k-1})}$. In this case cross terms look
like:
\begin{eqnarray*}
\Delta {\cal L} &=& i d_{k,k} \left\{
\phantom{|}^{\mu\nu\alpha\beta}_{abcd}
\right\} [ \bar{R}_{\mu\nu}{}^{(a_k),a(b_{k-1})} \Gamma^{bcd}
\Psi_{\alpha\beta}{}^{(a_k),(b_{k-1})} + 6k
\bar{R}_{\mu\nu}{}^{a(a_{k-1}),b(b_{k-1})} \gamma^c 
\Psi_{\alpha\beta}{}^{d(a_{k-1}),(b_{k-1})} \nonumber \\
 && \qquad \qquad  - \bar{\Psi}_{\mu\nu}{}^{(a_k),(a_{k-1})}
\Gamma^{abc} R_{\alpha\beta}{}^{(a_k),d(b_{k-1})} + 6k
\bar{\Psi}_{\mu\nu}{}^{a(a_{k-1}),(b_{k-1})} \gamma^b
R_{\alpha\beta}{}^{c(a_{k-1}),d(b_{k-1})} ]
\end{eqnarray*}
and to compensate for their non-invariance we have to introduce:
\begin{eqnarray}
\delta' R_{\mu\nu}{}^{(a_k),(b_k)} &=& - \frac{d_{k,k}}{20k(d+k-4)} [ 
\xi_{[\mu}{}^{(a_k),(b_{k-1}} e_{\nu]}{}^{b_1)} -
e_{[\nu}{}^{(a_1} \xi_{\mu]}{}^{a_{k-1})(b_1,b_{k-1})} + \dots ] 
\nonumber \\
\delta' \Psi_{\mu\nu}{}^{(a_k),(b_{k-1})} &=& - \frac{d_{k,k}}{10k}
\xi_{[\mu}{}^{(a_k),(b_{k-1})}{}_{\nu]}
\end{eqnarray}
where again dots stand for the additional terms which are necessary
for variations to be $\gamma$-transverse.

$\Psi_{\mu\nu}{}^{(a_k),(b_l)} \Leftrightarrow 
\Psi_{\mu\nu}{}^{(a_k),(b_{l-1})}$. Corresponding cross terms have
the following form:
\begin{eqnarray*}
(-1)^{k+l} \Delta {\cal L} &=&  i d_{k,l} 
\left\{ \phantom{|}^{\mu\nu\alpha\beta}_{abcd} \right\} [
\bar{\Psi}_{\mu\nu}{}^{(a_k),a(b_{l-1})} \Gamma^{bcd}
\Psi_{\alpha\beta}{}^{(a_k),(b_{l-1})} + \nonumber \\
 && \qquad \qquad + 6k
\bar{\Psi}_{\mu\nu}{}^{a(a_{k-1}),b(b_{l-1})} \gamma^c 
\Psi_{\alpha\beta}{}^{d(a_{k-1}),(b_{l-1})}  - \\
 && \qquad \qquad -
\bar{\Psi}_{\mu\nu}{}^{(a_k),(b_{l-1})} \Gamma^{abc}
\Psi_{\alpha\beta}{}^{(a_k),d(b_{l-1})} + \nonumber \\
 && \qquad \qquad  + 6k
\bar{\Psi}_{\mu\nu}{}^{a(a_{k-1}),(b_{l-1})} \gamma^b 
\Psi_{\alpha\beta}{}^{c(a_{k-1}),d(b_{l-1})} ]
\end{eqnarray*}
and to compensate for their non-invariance we have to introduce the
following corrections:
\begin{eqnarray}
\delta' \Psi_{\mu\nu}{}^{(a_k),(b_l)} &=& - 
\frac{d_{k,l}}{10l(k-l+2)(d+l-4)} [ (k-l+1) \xi_\mu{}^{(a_k),(b_{l-1}}
e_{\nu]}{}^{b_1)} - \nonumber \\
 && \qquad \qquad \qquad \qquad \qquad - e_{[\nu}{}^{(a_1} 
\xi_{\mu]}{}^{a_{k-1})(b_1,b_{l-1})} + \dots ] \nonumber \\
\delta' \Psi_{\mu\nu}{}^{(a_k),(b_{l-1})} &=& - \frac{d_{k,l}}{10l}
\xi_{[\mu}{}^{(a_k),(b_{l-1})}{}_{\nu]}
\end{eqnarray}

$\Psi_{\mu\nu}{}^{(a_k)} \Leftrightarrow \Phi_\mu{}^{(a_k)}$. This
case we have already considered in Subsection 3.2, so we will not
repeat corresponding formulas here.

Collecting all pieces together we obtain the following complete set of
gauge transformations:
\begin{eqnarray}
\delta R_{\mu\nu}{}^{(a_k),(b_k)} &=& D_{[\mu} 
\xi_{\nu]}{}^{(a_k),(b_k)} + \frac{i a_{k,k}}{5(d-4)} [ \gamma_{[\mu}
\xi_{\nu]}{}^{(a_k),(b_l)} + \dots] - \nonumber \\
 && - \frac{d_{k,k}}{20k(d+k-4)} [ 
\xi_{[\mu}{}^{(a_k),(b_{k-1}} e_{\nu]}{}^{b_1)} -
e_{[\nu}{}^{(a_1} \xi_{\mu]}{}^{a_{k-1})(b_1,b_{k-1})} + \dots ]
\nonumber \\
\delta \Psi_{\mu\nu}{}^{(a_k),(b_l)} &=& D_{[\mu}
\xi_{\nu]}{}^{(a_k),(b_l)} + \frac{i a_{k,l}}{5(d-4)} [ \gamma_{[\mu}
\xi_{\nu]}{}^{(a_k),(b_l)} + \dots ] - \nonumber \\
 && - \frac{d_{k,l}}{10l(k-l+2)(d+l-4)} [ (k-l+1) 
\xi_\mu{}^{(a_k),(b_{l-1}} e_{\nu]}{}^{b_1)} -  \\
 && - e_{[\nu}{}^{(a_1} \xi_{\mu]}{}^{a_{k-1})(b_1,b_{l-1})} + \dots ]
- \frac{d_{k,l+1}}{10(l+1)} \xi_{[\mu}{}^{(a_k),(b_l)}{}_{\nu]},
\qquad 1 \le l \le k-1 \nonumber \\
\delta \Psi_{\mu\nu}{}^{(a_k)} &=& D_{[\mu} \xi_{\nu]}{}^{(a_k)} +
\frac{i a_{k,0}}{5(d-4)} [ \gamma_{[\mu} \xi_{\nu]}{}^{(a_k)} + \dots]
- \frac{d_{k,1}}{10} \xi_{[\mu}{}^{(a_k),}{}_{\nu]} - \nonumber \\
 && - \frac{(k+1) d_{k,0}}{10(k+2)(d-3)(d-4)} [ \Gamma_{\mu\nu}
\zeta^{(a_k)} + \dots] \nonumber \\
\delta \Phi_\mu{}^{(a_k)} &=& D_\mu \zeta^{(a_k)} - 
\frac{i b_k}{3(d-2)} [ \gamma_\mu \zeta^{(a_k)} + \dots] + 2 d_{k,0}
\xi_\mu{}^{(a_k)} \nonumber
\end{eqnarray}

At this point we have complete Lagrangian as well as complete set of
gauge transformations, in this all parameters in gauge transformations
are expressed in terms of Lagrangian parameters $a_{k,l}$, $d_{k,l}$
and $b_k$ so that all variations with one derivative cancel. Now we
have to calculate all variations without derivatives (including
contributions of kinetic terms due to non-commutativity of covariant
derivatives) and require their cancellation. Once again we omit these
lengthy but straightforward calculations and give final results only.
First of all we obtain a number of recurrent relations on diagonal
mass terms $a_{k,l}$ which allow us to express all of them in terms of
the main one $a_{k,k} = M$:
$$
a_{k,l} = \frac{d+2k-2)}{(d+2l-2)} M, \qquad
b_k = - \frac{3(d+2k-2)}{5(d-4)} M
$$
Then we obtain recurrent relations on parameters $d_{k,l}$ which allow
us to express all of them in terms of the main one:
$$
d_{k,l}{}^2 = \frac{l(k-l+2)(d+k+l-3)}{(d+2l-4)}
[ m^2 - 100(k-l)(d+k+l-4) \kappa ], \qquad 1 \le l \le k
$$
$$
d_{k,0}{}^2 = \frac{(k+2)(d+k-3)}{10(d-4)}
[ m^2 - 100 k(d+k-4) \kappa] 
$$
where we introduced a notation:
$$
m^2 = \frac{(d+2k-4)}{2k(d+2k-3)} d_{k,k}{}^2
$$
At last we obtain an important relation on parameters $M$ and $m$:
$$
4 M^2 = m^2 - 25 (d+2k-4)^2 \kappa
$$

Let us analyze the results obtained. We have already seen in
Subsection 2.5 that massless spin-tensor $R_{\mu\nu}{}^{(a_k),(b_k)}$
admits deformation into $AdS$ space without introduction of any
additional fields. And indeed, as the last relation clearly shows, in
$AdS$ space ($\kappa < 0$) nothing prevent us from considering a limit
$m \to 0$ when all Goldstone fields decouple from the main one. From
the other hand, in the $dS$ space we again obtain unitary forbidden
region $m^2 < 25(d+2k-4)^2 \kappa$. At the boundary of this region all
diagonal mass terms become equal to zero so that the theory greatly
simplifies (though the number of physical degrees of freedom remains
to be the same). Inside forbidden region we find a number of 
(non-unitary) partially massless theories. They appear each time when
one of the parameters $d_{k,l} \to 0$. In this, the whole system
decomposes into two disconnected subsystems containing the fields
$R_{\mu\nu}{}^{(a_k),(b_l)}$, $\Psi_{\mu\nu}{}^{(a_k),(b_n)}$ $l \le n
\le k-1$ and $\Psi_{\mu\nu}{}^{(a_k),(b_n)}$ $0 \le n \le l-1$,
$\Phi_\mu{}^{(a_k)}$, correspondingly. 

\subsection{$Y(k+\frac{3}{2},l+\frac{3}{2})$} 

Now we are ready to consider general case --- massive spin-tensor
$Y(k+\frac{3}{2},l+\frac{3}{2})$ with $k > l \ge 1$. Our usual
procedure (consider gauge transformations for all fields and take into
account their reducibility) leads to the following set of fields which
are necessary for gauge invariant description: 
$Y(m+\frac{3}{2},n+\frac{3}{2})$ and $Y(m+\frac{3}{2},\frac{1}{2})$
where $l \le m \le k$ and $0 \le n \le l$. These fields as well as
parameters determining appropriate cross terms (see below) are shown
on Figure 7.
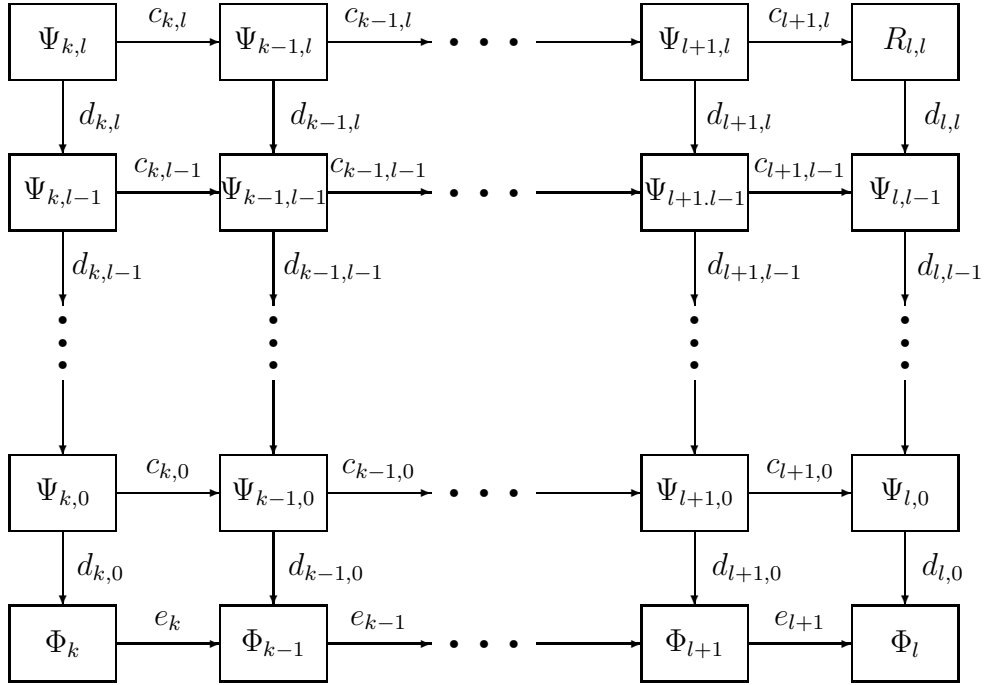
\begin{figure}[htb]
\begin{center}
\begin{picture}(146,92)
\put(10,1){\framebox(14,10)[]{$\Phi_k$}}
\put(27,6){\makebox(8,6)[]{$e_{k}$}}
\put(38,1){\framebox(14,10)[]{$\Phi_{k-1}$}}
\put(55,6){\makebox(8,6)[]{$e_{k-1}$}}
\put(94,1){\framebox(14,10)[]{$\Phi_{l+1}$}}
\put(111,6){\makebox(8,6)[]{$e_{l+1}$}}
\put(122,1){\framebox(14,10)[]{$\Phi_l$}}
\multiput(24,6)(28,0){4}{\vector(1,0){14}}
\multiput(69,6)(4,0){3}{\circle*{1}}
\put(10,21){\framebox(14,10)[]{$\Psi_{k,0}$}}
\put(27,26){\makebox(8,6)[]{$c_{k,0}$}}
\put(18,13){\makebox(8,6)[]{$d_{k,0}$}}
\put(38,21){\framebox(14,10)[]{$\Psi_{k-1,0}$}}
\put(55,26){\makebox(8,6)[]{$c_{k-1,0}$}}
\put(48,13){\makebox(8,6)[]{$d_{k-1,0}$}}
\put(94,21){\framebox(14,10)[]{$\Psi_{l+1,0}$}}
\put(111,26){\makebox(8,6)[]{$c_{l+1,0}$}}
\put(104,13){\makebox(8,6)[]{$d_{l+1,0}$}}
\put(122,21){\framebox(14,10)[]{$\Psi_{l,0}$}}
\put(130,13){\makebox(8,6)[]{$d_{l,0}$}}
\multiput(24,26)(28,0){4}{\vector(1,0){14}}
\multiput(69,26)(4,0){3}{\circle*{1}}
\put(10,61){\framebox(14,10)[]{$\Psi_{k,l-1}$}}
\put(27,66){\makebox(8,6)[]{$c_{k,l-1}$}}
\put(19,53){\makebox(8,6)[]{$d_{k,l-1}$}}
\put(38,61){\framebox(14,10)[]{$\Psi_{k-1,l-1}$}}
\put(55,66){\makebox(8,6)[]{$c_{k-1,l-1}$}}
\put(49,53){\makebox(8,6)[]{$d_{k-1,l-1}$}}
\put(94,61){\framebox(14,10)[]{$\Psi_{l+1.l-1}$}}
\put(111,66){\makebox(8,6)[]{$c_{l+1,l-1}$}}
\put(105,53){\makebox(8,6)[]{$d_{l+1,l-1}$}}
\put(122,61){\framebox(14,10)[]{$\Psi_{l,l-1}$}}
\put(131,53){\makebox(8,6)[]{$d_{l,l-1}$}}
\multiput(24,66)(28,0){4}{\vector(1,0){14}}
\multiput(69,66)(4,0){3}{\circle*{1}}
\put(10,81){\framebox(14,10)[]{$\Psi_{k,l}$}}
\put(27,86){\makebox(8,6)[]{$c_{k,l}$}}
\put(18,73){\makebox(8,6)[]{$d_{k,l}$}}
\put(38,81){\framebox(14,10)[]{$\Psi_{k-1,l}$}}
\put(55,86){\makebox(8,6)[]{$c_{k-1,l}$}}
\put(48,73){\makebox(8,6)[]{$d_{k-1,l}$}}
\put(94,81){\framebox(14,10)[]{$\Psi_{l+1,l}$}}
\put(111,86){\makebox(8,6)[]{$c_{l+1,l}$}}
\put(103,73){\makebox(8,6)[]{$d_{l+1,l}$}}
\put(122,81){\framebox(14,10)[]{$R_{l,l}$}}
\put(130,73){\makebox(8,6)[]{$d_{l,l}$}}
\multiput(24,86)(28,0){4}{\vector(1,0){14}}
\multiput(69,86)(4,0){3}{\circle*{1}}
\multiput(17,21)(0,20){4}{\vector(0,-1){10}}
\multiput(17,43)(0,3){3}{\circle*{1}}
\multiput(45,21)(0,20){4}{\vector(0,-1){10}}
\multiput(45,43)(0,3){3}{\circle*{1}}
\multiput(101,21)(0,20){4}{\vector(0,-1){10}}
\multiput(101,43)(0,3){3}{\circle*{1}}
\multiput(129,21)(0,20){4}{\vector(0,-1){10}}
\multiput(129,43)(0,3){3}{\circle*{1}}
\end{picture}
\end{center}
\caption{General massive $Y(k+\frac{3}{2},l+\frac{3}{2})$ theory}
\end{figure}

As in all previous cases, the total Lagrangian contains kinetic and
diagonal mass terms for all fields as well as cross terms without
derivatives:
\begin{equation}
{\cal L} = \sum_{m=l}^k \sum_{n=0}^l  {\cal L}
(\Psi_{\mu\nu}{}^{(a_m),(b_n)}) + \sum_{m=l}^{k} {\cal L}
(\Phi_\mu{}^{(a_m)}) + {\cal L}_{cross}
\end{equation}
Recall that cross terms appear for nearest neighbours only, i.e. main
gauge field with primary fields, primary with secondary ones and so
on. Thus general $\Psi_{\mu\nu}{}^{(a_m),(b_n)}$ field has cross terms
with four other fields as shown on Figure 8.
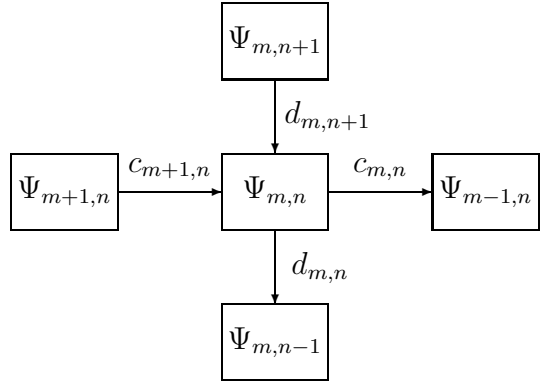
\begin{figure}[htb]
\begin{center}
\begin{picture}(90,52)
\put(38,1){\framebox(14,10)[]{$\Psi_{m,n-1}$}}
\put(10,21){\framebox(14,10)[]{$\Psi_{m+1,n}$}}
\put(24,26){\vector(1,0){14}}
\put(27,26){\makebox(8,6)[]{$c_{m+1,n}$}}
\put(38,21){\framebox(14,10)[]{$\Psi_{m,n}$}}
\put(52,26){\vector(1,0){14}}
\put(55,26){\makebox(8,6)[]{$c_{m,n}$}}
\put(45,21){\vector(0,-1){10}}
\put(47,13){\makebox(8,6)[]{$d_{m,n}$}}
\put(66,21){\framebox(14,10)[]{$\Psi_{m-1,n}$}}
\put(38,41){\framebox(14,10)[]{$\Psi_{m,n+1}$}}
\put(45,41){\vector(0,-1){10}}
\put(48,33){\makebox(8,6)[]{$d_{m,n+1}$}}
\end{picture}
\end{center}
\caption{Illustration on possible cross terms}
\end{figure}
 In the previous Subsection we have already considered cross terms for
the pair $\Psi_{\mu\nu}{}^{(a_m),(b_n)} \Leftrightarrow
\Psi_{\mu\nu}{}^{(a_m),(b_{n-1})}$, thus the only new terms we need
are cross terms for the pair $\Psi_{\mu\nu}{}^{(a_m),(b_n)}
\Leftrightarrow \Psi_{\mu\nu}{}^{(a_{m-1}),(b_n)}$. They look as
follows:
\begin{eqnarray*}
(-1)^{m+n} \Delta {\cal L} &=& i  c_{m,n} \left\{
\phantom{|}^{\mu\nu\alpha\beta}_{abcd} \right\} [
\bar{\Psi}{}_{\mu\nu}{}^{a(a_{m-1}),(b_n)} \Gamma^{bcd}
\Psi_{\alpha\beta}{}^{(a_{m-1}),(b_n)} + \\
 && \qquad \qquad \qquad - 6n
\bar{\Psi}_{\mu\nu}{}^{a(a_{m-1}),b(b_{n-1})} \gamma^c
\Psi_{\alpha\beta}{}^{(a_{n-1}),d(b_{n-1})} - \\
 && \qquad \qquad \qquad - \bar{\Psi}_{\mu\nu}{}^{(a_{m-1}),(b_n)} 
\Gamma^{abc} \Psi_{\alpha\beta}{}^{d(a_{m-1}),(b_n)} + \\
 && \qquad \qquad \qquad - 6n
\bar{\Psi}_{\mu\nu}{}^{(a_{m-1}),a(b_{n-1})} \gamma^b
\Psi_{\alpha\beta}{}^{c(a_{m-1}),d(b_{n-1})} ]
\end{eqnarray*}
In this, to compensate for their non-invariance we have to introduce
the following corrections to gauge transformations:
\begin{eqnarray}
\delta' \Psi_{\mu\nu}{}^{(a_m),(b_n)} &=& \frac{c_{m,n}}{10m(d+m-3)}
[ e_{[\mu}{}^{(a_1} \xi_{\nu]}{}^{a_{m-1}),(b_n)} + \dots ] \\
\delta' \Psi_{\mu\nu}{}^{(a_{m-1}),(b_n)} &=& -
\frac{c_{m,n}}{10(m-n+1)(m+1)} [ (m-n+1) 
\xi_{[\mu,\nu]}{}^{(a_{m-1}),(b_n)} + 
\xi_{[\mu}{}^{(a_{m-1})(b_1,b_{n-1})}{}_{\nu]} ] \nonumber
\end{eqnarray}
Note also that the last two rows on Figure 7 are to those 
on the Figure 4 in Subsection 3.2, so all necessary terms have
already been considered there.

Collecting all pieces together we obtain the following complete set of
cross terms:
\begin{eqnarray}
{\cal L}_{cross} &=& i \sum_{m=l}^k \sum_{n=1}^l (-1)^{m+n}
\left\{ \phantom{|}^{\mu\nu\alpha\beta}_{abcd} \right\} \left\{
c_{m,n} [
\bar{\Psi}{}_{\mu\nu}{}^{a(a_{m-1}),(b_n)} \Gamma^{bcd}
\Psi_{\alpha\beta}{}^{(a_{m-1}),(b_n)} + \right. \nonumber \\
 && \qquad \qquad \qquad \qquad - 6n
\bar{\Psi}_{\mu\nu}{}^{a(a_{m-1}),b(b_{n-1})} \gamma^c
\Psi_{\alpha\beta}{}^{(a_{n-1}),d(b_{n-1})} - \nonumber \\
 && \qquad \qquad \qquad \qquad - 
\bar{\Psi}_{\mu\nu}{}^{(a_{m-1}),(b_n)} 
\Gamma^{abc} \Psi_{\alpha\beta}{}^{d(a_{m-1}),(b_n)} + \nonumber \\
 && \qquad \qquad \qquad \qquad - 6n
\bar{\Psi}_{\mu\nu}{}^{(a_{m-1}),a(b_{n-1})} \gamma^b
\Psi_{\alpha\beta}{}^{c(a_{m-1}),d(b_{n-1})} ] + \nonumber \\
 && \qquad \qquad \qquad + d_{m,n}  [
\bar{\Psi}_{\mu\nu}{}^{(a_m),a(b_{n-1})} \Gamma^{bcd}
\Psi_{\alpha\beta}{}^{(a_m),(b_{n-1})} + \nonumber \\
 && \qquad \qquad \qquad \qquad + 6m
\bar{\Psi}_{\mu\nu}{}^{a(a_{m-1}),b(b_{n-1})} \gamma^c 
\Psi_{\alpha\beta}{}^{d(a_{m-1}),(b_{n-1})} - \nonumber \\
 && \qquad \qquad \qquad \qquad -
\bar{\Psi}_{\mu\nu}{}^{(a_m),(b_{n-1})} \Gamma^{abc}
\Psi_{\alpha\beta}{}^{(a_m),d(b_{n-1})} + \nonumber \\
 && \qquad \qquad \qquad \qquad \left. + 6m
\bar{\Psi}_{\mu\nu}{}^{a(_{m-1}),(b_{n-1})} \gamma^b 
\Psi_{\alpha\beta}{}^{c(a_{m-1}),d(b_{n-1})} ] \right\} \nonumber \\
 && + i \sum_{m=l}^k (-1)^m 
\left\{ \phantom{|}^{\mu\nu\alpha\beta}_{abcd} \right\} \left\{
c_{m,0} [
\bar{\Psi}_{\mu\nu}{}^{a(a_m)} \Gamma^{bcd} 
\Psi_{\alpha\beta}{}^{(a_m)} - \bar{\Psi}_{\mu\nu}{}^{(a_m)}
\Gamma^{abc} \Psi_{\alpha\beta}{}^{d(a_m)} ] + \right. \nonumber \\
 && \qquad \qquad \qquad + d_{m,0}  [
\bar{\Psi}_{\mu\nu}{}^{(a_m)} \Gamma^{abc} \Phi_\alpha{}^{(a_m)} -
6m \bar{\Psi}_{\mu\nu}{}^{a(a_{m-1})} \gamma^b 
\Phi_\alpha{}^{c(a_{m-1})}  + \nonumber \\
 && \qquad \qquad \qquad \qquad + 
\bar{\Psi}_{\mu\nu}{}^{(a_m)} \Gamma^{abc} \Phi_\alpha{}^{(a_m)} -
6m \bar{\Psi}_{\mu\nu}{}^{a(a_{m-1})} \gamma^b 
\Phi_\alpha{}^{c(a_{m-1})} ] + \nonumber \\
 && \left. \qquad \qquad \qquad + e_m  [
\bar{\Phi}_\mu{}^{a(a_m)} \gamma^b \Phi_\nu{}^{(a_m)} -
\bar{\Phi}_\mu{}^{(a_m)} \gamma^a \Phi_\nu{}^{b(a_m)} ] \right\}
\end{eqnarray}
Similarly, combining results of this and previous Subsections, we
obtain complete set of gauge transformations for all fields involved:
\begin{eqnarray}
\delta \Psi_{\mu\nu}{}^{(a_m),(b_n)} &=& D_{[\mu}
\xi_{\nu]}{}^{(a_m),(b_n)} + \frac{i a_{m,n}}{5(d-4)} [ \gamma_{[\mu}
\xi_{\nu]}{}^{(a_m),(b_n)} + \dots]  - \nonumber \\
 && - \frac{c_{m+1,n}}{10(m-n+2)(m+2)} [ (m-n+2)
\xi_{[\mu,\nu]}{}^{(a_m),(b_n)} + 
\xi_{[\mu}{}^{(a_m)(b_1,b_{n-1})}{}_{\nu]} ] - \nonumber \\
 && - \frac{c_{m,n}}{10m(d+m-3)} [ e_{[\mu}{}^{(a_1} 
\xi_{\nu]}{}^{a_{m-1}),(b_n)} + \dots ] - \frac{d_{m,n+1}}{10(n+1)}
\xi_{[\mu}{}^{(a_m),(b_{n-1})}{}_{\nu]} - \nonumber \\
 && - \frac{d_{m,n}}{10n(m-n+2)(d+n-4)} [ (m-n+1)
\xi_{[\mu}{}^{(a_m),(b_{n-1}} e_{\nu]}{}^{b_1)} - \nonumber \\
 && \qquad \qquad \qquad \qquad \qquad \qquad -
e_{[\nu}{}^{(a_1} \xi_{\mu]}{}^{a_{m-1})(b_1,b_{n-1})} + \dots ],
\qquad 1 \le n \le l \nonumber \\
\delta \Psi_{\mu\nu}{}^{(a_m)} &=& D_{[\mu} \xi_{\nu]}{}^{(a_m)} +
\frac{i a_{m,0}}{5(d-4)} [ \gamma_{[\mu} \xi_{\nu]}{}^{(a_m)} + \dots]
+ \frac{c_{m,0}}{10(m+2)} \xi_{[\mu,\nu]}{}^{(a_m)} -  \\
 && - \frac{c_{m-1,0}}{10l(d+m-3)} [ e_{[\mu}{}^{(a_1} 
\xi_{\nu]}{}^{a_{m-1})} + \dots] - \nonumber \\
 && - \frac{(m+1) d_{m,0}}{10(m+2)(d-3)(d-5)}
[ \Gamma_{\mu\nu} \zeta^{(a_m)} + \dots] \nonumber \\
\delta \Phi_\mu{}^{(a_m)} &=& D_\mu \zeta^{(a_m)} - 
\frac{i b_m}{3(d-2)} [ \gamma_\mu \zeta^{(a_m)} + \dots] + 2 d_{m,0}
\xi_\mu{}^{(a_m)} + \frac{e_m}{6(m+1)} \zeta^{(a_m)} + \nonumber \\
 && + \frac{e_{m-1}}{6m(d+m-2)} [ e_\mu{}^{(a_1} \zeta^{a_{m-1})} +
\dots ] \nonumber
\end{eqnarray}

Having in our disposal total Lagrangian and complete set of gauge
transformations where all variations with one derivative cancel, we
proceed with variations without derivatives. After lengthy but
straightforward calculations we obtain the following results.

First of all we obtain a number of relations on diagonal mass terms
$a_{m,n}$ and $b_m$ which allow us to express all them in terms of
main one $a_{k,l} = M$:
$$
a_{m,n} = \frac{(d+2k)(d+2l-2)}{(d+2m)(d+2n-2)} M
$$
Similarly, we get a number of relations on the parameters $d_{m,n}$,
$c_{m,n}$ and $e_m$ (determining cross terms) so that all of them can
be expressed in terms of one main parameter. We choose $d_{k,l}$ as
such main parameter and introduce a notation:
$$
m^2 = \frac{(k-l+1)(d+2l-4)}{l(k-l+2)(d+2l-3)} d_{k,l}{}^2
$$
Then we obtain the following important expressions for the parameters
$d_{k,n}$ corresponding to leftmost column on Figure 7:
$$
d_{k,n}{}^2 = \frac{n(l-n+1)(k-n+2)(d+l+n-3)}{(k-n+1)(d+2n-4)}
[ m^2 - 100 (l-n) (d+l+n-4) \kappa ], \quad n \ge 1
$$
$$
d_{k,0}{}^2 = \frac{(k+2)(l+1)(d+l-3)}{10(k+1)(d-4)}
[ m^2 - 100 l (d+l-4) \kappa ]
$$
as well as for parameters $c_{m,l}$ corresponding to topmost row on
Figure 7:
$$
c_{m,l}{}^2 = \frac{m(k-m+1)(d+k+m-1)}{(d+2m-2)}
[ m^2 + 100 (m-l+1)(d+m+l-3) \kappa ]
$$
It is very important (and this gives a nice check for all
calculations) that all parameters $d_{m,n}$ corresponding to the same
row on Figure 7 turn out to be proportional to the leftmost one
$d_{k,n}$:
$$
d_{m,n}{}^2 = \frac{(k-n+1)(d+k+n-2)}{(m-n+1)(d+m+n-2)} d_{k,n}{}^2
$$
Similarly, all parameters $c_{m,n}$ and $e_m$ corresponding to the
same column turn out to be proportional to the topmost one $c_{m,l}$:
$$
c_{m,n}{}^2 = \frac{(m-l)(d+m+l-2)}{(m-n)(d+m+n-2)} c_{m,l}{}^2
$$
$$
e_m{}^2 = \frac{9(m-l)(d+m+l-2)}{25(m+1)(d+m-3)} c_{m,l}{}^2
$$
At last but not least, we obtain an important relation on two main
parameters $M$ and $m$:
$$
4 M^2 = m^2 - 25 (d+2l-4)^2 \kappa
$$

We have already mentioned in Subsection 2.6 that massless spin-tensor 
$\Psi_{\mu\nu}{}^{(a_k),(b_l)}$ does not admit deformation into $AdS$
space without introduction of additional fields. In the gauge
invariant formulation for massive spin-tensor such a limit would
require that both $d_{k,l} \to 0$ and $c_{k,l} \to 0$ simultaneously
and such possibility exists in flat Minkowski space ($\kappa = 0$)
only. For non-zero values of cosmological constant we obtain a number
of partially massless limits instead.

Let us consider $AdS$ space ($\kappa < 0$) first. The most physically
interesting limit arises when $m^2 = - 100(k-l+1)(d+k+l-3) \kappa$. In
this, parameter $c_{k,l}$ (and hence all parameters $c_{k,n}$ and
$e_k$) becomes equals to zero and fields
$\Psi_{\mu\nu}{}^{(a_k),(b_n)}$ $0 \le n \le l$ and
$\Phi_\mu{}^{(a_k)}$ (corresponding to leftmost column in Figure 7)
decouple and describe (the only) unitary partially massless theory.
The Lagrangian for this theory has the form:
\begin{equation}
{\cal L} = \sum_{n=0}^l  {\cal L} (\Psi_{\mu\nu}{}^{(a_k),(b_n)}) +
{\cal L} (\Phi_\mu{}^{(a_k)}) + {\cal L}_{cross}
\end{equation}
\begin{eqnarray}
{\cal L}_{cross} &=& i \sum_{n=1}^l (-1)^{k+n} d_{k,n}
\left\{ \phantom{|}^{\mu\nu\alpha\beta}_{abcd} \right\} [
\bar{\Psi}_{\mu\nu}{}^{(a_k),a(b_{n-1})} \Gamma^{bcd}
\Psi_{\alpha\beta}{}^{(a_k),(b_{n-1})} +  \nonumber \\
 && \qquad \qquad \qquad \qquad + 6k
\bar{\Psi}_{\mu\nu}{}^{a(a_{k-1}),b(b_{n-1})} \gamma^c 
\Psi_{\alpha\beta}{}^{d(a_{k-1}),(b_{n-1})} - \nonumber \\
 && \qquad \qquad \qquad \qquad -
\bar{\Psi}_{\mu\nu}{}^{(a_k),(b_{n-1})} \Gamma^{abc}
\Psi_{\alpha\beta}{}^{(a_k),d(b_{n-1})} + \nonumber \\
 && \qquad \qquad \qquad \qquad + 6k
\bar{\Psi}_{\mu\nu}{}^{a(_{k-1}),(b_{n-1})} \gamma^b 
\Psi_{\alpha\beta}{}^{c(a_{k-1}),d(b_{n-1})} ] \nonumber \\
 && + i d_{k,0}
\left\{ \phantom{|}^{\mu\nu\alpha\beta}_{abcd} \right\} [
\bar{\Psi}_{\mu\nu}{}^{(a_k)} \Gamma^{abc} \Phi_\alpha{}^{(a_k)} -
6k \bar{\Psi}_{\mu\nu}{}^{a(a_{k-1})} \gamma^b 
\Phi_\alpha{}^{c(a_{k-1})}  + \nonumber \\
 && \qquad \qquad \qquad + 
\bar{\Psi}_{\mu\nu}{}^{(a_k)} \Gamma^{abc} \Phi_\alpha{}^{(a_k)} -
6k \bar{\Psi}_{\mu\nu}{}^{a(a_{k-1})} \gamma^b 
\Phi_\alpha{}^{c(a_{k-1})} ]
\end{eqnarray}
and is invariant under the following gauge transformations:
\begin{eqnarray}
\delta \Psi_{\mu\nu}{}^{(a_k),(b_n)} &=& D_{[\mu}
\xi_{\nu]}{}^{(a_k),(b_n)} + \frac{i a_{k,n}}{5(d-4)} [ \gamma_{[\mu}
\xi_{\nu]}{}^{(a_k),(b_n)} + \dots] 
- \frac{d_{k,n+1}}{10(n+1)}
\xi_{[\mu}{}^{(a_k),(b_{n-1})}{}_{\nu]} - \nonumber \\
 && - \frac{d_{k,n}}{10n(k-n+2)(d+n-4)} [ (k-n+1)
\xi_{[\mu}{}^{(a_k),(b_{n-1}} e_{\nu]}{}^{b_1)} - \nonumber \\
 && \qquad \qquad \qquad \qquad \qquad \qquad -
e_{[\nu}{}^{(a_1} \xi_{\mu]}{}^{a_{k-1})(b_1,b_{n-1})} + \dots ],
\qquad 1 \le n \le l \nonumber \\
\delta \Psi_{\mu\nu}{}^{(a_k)} &=& D_{[\mu} \xi_{\nu]}{}^{(a_k)} +
\frac{i a_{k,0}}{5(d-4)} [ \gamma_{[\mu} \xi_{\nu]}{}^{(a_k)} + \dots]
 -  \\
 && - \frac{(k+1) d_{k,0}}{10(k+2)(d-3)(d-5)}
[ \Gamma_{\mu\nu} \zeta^{(a_k)} + \dots] \nonumber \\
\delta \Phi_\mu{}^{(a_k)} &=& D_\mu \zeta^{(a_k)} - 
\frac{i b_k}{3(d-2)} [ \gamma_\mu \zeta^{(a_k)} + \dots] + 2 d_{k,0}
\xi_\mu{}^{(a_k)} \nonumber
\end{eqnarray}
 All other fields just give massive theory for the
spin-tensor $\Psi_{\mu\nu}{}^{(a_{k-1}),(b_l)}$. Besides, a number of
non-unitary partially massless limits exist. It happens each time then
one of the $c_{m,l}$ (and hence all $c_{m,n}$ with $0 \le n \le l$ and
$e_m$) goes to zero. In this, the whole system decomposes into two
disconnected subsystems (and diagram on Figure 7 splits horizontally
into two blocks as shown on Figure 9).
\begin{figure}[htb]
\begin{center}
\begin{picture}(174,52)
\put(10,1){\framebox(14,10)[]{$\Phi_k$}}
\put(17,21){\vector(0,-1){10}}
\multiput(17,23)(0,3){3}{\circle*{1}}
\put(17,41){\vector(0,-1){10}}
\put(10,41){\framebox(14,10)[]{$\Psi_{k,l}$}}
\put(24,6){\vector(1,0){14}}
\multiput(41,6)(4,0){3}{\circle*{1}}
\put(52,6){\vector(1,0){14}}
\put(66,1){\framebox(14,10)[]{$\Phi_m$}}
\put(24,46){\vector(1,0){14}}
\multiput(41,46)(4,0){3}{\circle*{1}}
\put(52,46){\vector(1,0){14}}
\put(66,41){\framebox(14,10)[]{$\Psi_{m,l}$}}
\put(73,21){\vector(0,-1){10}}
\multiput(73,23)(0,3){3}{\circle*{1}}
\put(73,41){\vector(0,-1){10}}

\put(94,1){\framebox(14,10)[]{$\Phi_{m-1}$}}
\put(101,21){\vector(0,-1){10}}
\multiput(101,23)(0,3){3}{\circle*{1}}
\put(101,41){\vector(0,-1){10}}
\put(94,41){\framebox(14,10)[]{$\Psi_{m-1,l}$}}
\put(108,6){\vector(1,0){14}}
\multiput(125,6)(4,0){3}{\circle*{1}}
\put(136,6){\vector(1,0){14}}
\put(150,1){\framebox(14,10)[]{$\Phi_l$}}
\put(108,46){\vector(1,0){14}}
\multiput(125,46)(4,0){3}{\circle*{1}}
\put(136,46){\vector(1,0){14}}
\put(150,41){\framebox(14,10)[]{$R_{l,l}$}}
\put(157,21){\vector(0,-1){10}}
\multiput(157,23)(0,3){3}{\circle*{1}}
\put(157,41){\vector(0,-1){10}}
\end{picture}
\end{center}
\caption{Example of non-unitary partially massless limit in $AdS$}
\end{figure}
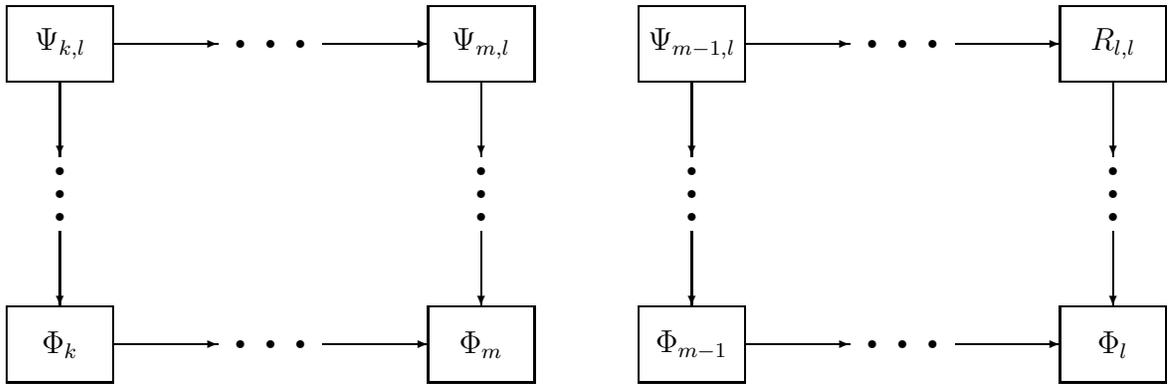
The left block describes a non-unitary partially massless theory,
while the right one gives massive theory for spin-tensor 
$\Psi_{\mu\nu}{}^{(a_{m-1}),(b_l)}$.  Recall that our definition of
masslessness is bounded to flat Minkowski space. From the anti de
Sitter group point of view each vertical column on Figure 7
corresponds to unitary irreducible representation which can be called
massless \cite{BMV00}. In this, all other representations (massive or
partially massless) can be constructed out of appropriate set of
massless ones as it should be.

Let us turn to the $dS$ space ($\kappa > 0$). Here we once again face
an unitary forbidden region $m^2 < 25(d+2l-4)^2 \kappa$ (which follows
from the relation between $M$ and $m$). Inside this forbidden region
we obtain a number of partially massless limits (but all of them lead
to the non-unitary theories). The first one arises then parameter
$d_{k,l}$ (and hence all parameters $d_{k,n}$) becomes zero. In this,
the fields $\Psi_{\mu\nu}{}^{(a_m),(b_l)}$ with $l \le m \le k$
(corresponding to upper row on Figure 7, see Figure 10) decouple and
describe partially massless theory which corresponds to irreducible
representation of the de Sitter group (and from the de Sitter group
point of view can be called massless). 
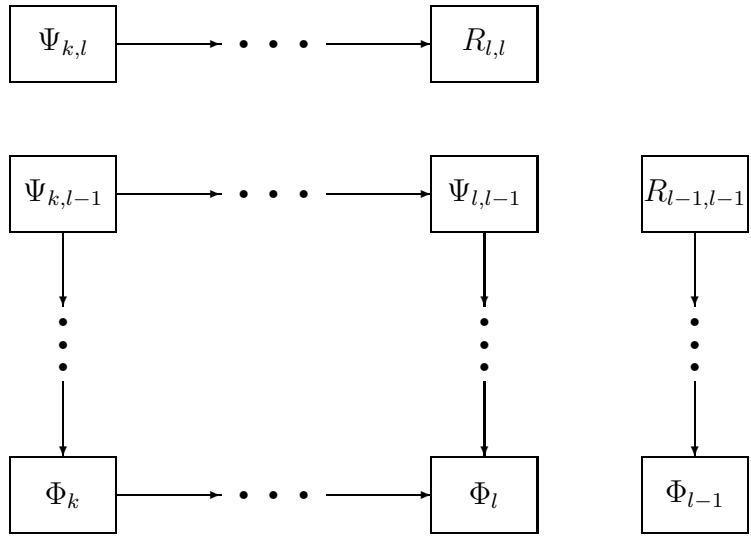
\begin{figure}[htb]
\begin{center}
\begin{picture}(118,72)
\put(10,1){\framebox(14,10)[]{$\Phi_k$}}
\put(17,21){\vector(0,-1){10}}
\multiput(17,23)(0,3){3}{\circle*{1}}
\put(17,41){\vector(0,-1){10}}
\put(10,41){\framebox(14,10)[]{$\Psi_{k,l-1}$}}
\put(24,6){\vector(1,0){14}}
\multiput(41,6)(4,0){3}{\circle*{1}}
\put(52,6){\vector(1,0){14}}
\put(66,1){\framebox(14,10)[]{$\Phi_l$}}
\put(24,46){\vector(1,0){14}}
\multiput(41,46)(4,0){3}{\circle*{1}}
\put(52,46){\vector(1,0){14}}
\put(66,41){\framebox(14,10)[]{$\Psi_{l,l-1}$}}
\put(73,21){\vector(0,-1){10}}
\multiput(73,23)(0,3){3}{\circle*{1}}
\put(73,41){\vector(0,-1){10}}
\put(10,61){\framebox(14,10)[]{$\Psi_{k,l}$}}
\put(24,66){\vector(1,0){14}}
\multiput(41,66)(4,0){3}{\circle*{1}}
\put(52,66){\vector(1,0){14}}
\put(66,61){\framebox(14,10)[]{$R_{l,l}$}}
\put(94,1){\framebox(14,10)[]{$\Phi_{l-1}$}}
\put(94,41){\framebox(14,10)[]{$R_{l-1,l-1}$}}
\put(101,21){\vector(0,-1){10}}
\multiput(101,23)(0,3){3}{\circle*{1}}
\put(101,41){\vector(0,-1){10}}
\end{picture}
\end{center}
\caption{Partially massless limit in $dS$ space}
\end{figure}
Contrary to what we have seen in $AdS$ case, all other fields also
gives partially massless theory. The reason is that to describe
complete massive theory for spin-tensor 
$\Psi_{\mu\nu}{}^{(a_k),(b_{l-1})}$ we need one more column of fields
as also shown on Figure 10.

Similarly, partially massless limits happens each time when one of the
parameters $d_{k,n}$ (and hence all parameters $d_{m,n}$ with $l \le m
\le k$) becomes zero. Once again the whole system decomposes into two
disconnected subsystems (and diagram on Figure 7 splits vertically
into two blocks). In this, both upper and bottom blocks describe
non-unitary partially massless theories. The reason again is that
bottom block does not have enough fields for description of massive
spin-tensor $\Psi_{\mu\nu}{}^{(a_k),(b_{n-1})}$.

\section{Conclusion}

Once again we have seen that frame-like formalism gives a simple and
elegant way for description of (spin)-tensors with different symmetry
properties. The formulation for massless mixed symmetry spin-tensors
constructed here turns out to be natural and straightforward
generalization of Skvortsov formulation for massless mixed symmetry
tensors \cite{Skv08} as well as Vasiliev formulation for completely
symmetric spin-tensors \cite{LV88,Vas88}. Similarly, all results on
massive mixed symmetry spin-tensors obtained here appear as natural
extension of previous results on massive (spin)-tensors
\cite{Zin08b,Zin08c,Zin09a}.

As a byproduct of our investigations, we obtain a generalization of
the results \cite{BMV00} for anti de Sitter group to the case of de
Sitter one. Recall, that in \cite{BMV00} it was shown that massless
(from anti de Sitter group point of view) representations contain more
degrees of freedom then corresponding Minkowski one and in the flat
space limit decompose into sum of massless Minkowski fields. For the
(spin)-tensors corresponding to Young tableau with two rows a
necessary pattern of massless fields can be obtain by cutting boxes
from the second row until we end up with the tableau with one row
corresponding to completely symmetric (spin)-tensor. Similarly, we
have
seen in Subsection 3.5 that massless (from the de Sitter group point
of view) representations correspond to a number of massless Minkowski
ones, in this necessary pattern can be obtained by cutting boxes from
the first row until we end up with the rectangular tableau. Note that
in both cases the procedure stops at the field having one its own
gauge transformation only.

Let us stress once again that one of the nice features of gauge
invariant formulation for massive fields is that it nicely works both
in flat Minkowski space as well as in $(A)dS$ space with arbitrary
value of cosmological constant. In particular, this allows us to
investigate all possible special partially massless limits that exist
both in $AdS$ as well as in $dS$ spaces. As we have seen, most of
these partially massless theories turn out to non-unitary. Thus
besides general massive theories the most/only physically interesting 
cases correspond to massless (in anti de Sitter sense) fields in $AdS$
space, in this general massive theory can be considered as smooth
deformation for appropriate collection of such massless fields.
Recall that till now most of results on higher spin interactions (see
e.g. recent reviews \cite{Vas04,Sor04,BCIV05,FT08}) were obtained for
massless fields in $AdS$ space. Thus it seems very interesting and
important to understand how such interacting theories for massless
fields in $AdS$ space could be deformed into the ones for massive
fields in flat Minkowski space. Some first very modest but
nevertheless encouraging results in this direction were obtained
recently \cite{Zin08,Zin09}.

\end{document}